\documentclass[12pt]{article}

\usepackage{amssymb}
\usepackage{amsmath}
\usepackage{amscd}
\usepackage{latexsym}
\usepackage{graphicx}

\usepackage{cite}

\newcommand{\be}{\begin{equation}}
\newcommand{\ee}{\end{equation}}

\newcommand{\bt}{\beta}
\newcommand{\vp}{\varphi}

\newcommand{\al}{\alpha}
\newcommand{\ra}{\rightarrow}
\newcommand{\sgm}{\sigma}

\newcommand{\om}{\omega}

\newcommand{\lbd}{\lambda}

\begin{document}

\begin{center}

{\Large{\bf Self-Similar Bridge between Regular and Critical Regions} \\ [5mm]

Vyacheslav I. Yukalov$^{1,2,a}$, Elizaveta P. Yukalova$^{3}$, and \\
Didier Sornette$^{4,5}$} \\ [3mm]

{\it
$^1$Bogolubov Laboratory of Theoretical Physics, \\
Joint Institute for Nuclear Research, Dubna 141980, Russia \\ [3mm]

$^2$Instituto de Fisica de S\~ao Carlos, Universidade de S\~ao Paulo, \\
CP 369, S\~ao Carlos 13560-970, S\~ao Paulo, Brazil \\ [2mm]

$^3$Laboratory of Information Technologies, \\
Joint Institute for Nuclear Research, Dubna 141980, Russia \\ [3mm]

$^4$Institute of Risk Analysis, Prediction and Management (Risks-X), \\
Southern University of Science and Technology, Shenzhen, People's Republic of China\\ [3mm]

$^5$Swiss Finance Institute, c/o University of Geneva, \\
40 blvd. Du Pont d'Arve, CH 1211 Geneva 4, Switzerland}

\end{center}

\vskip 2cm

\begin{abstract}

In statistical and nonlinear systems, two qualitatively distinct parameter regions are 
typically identified: the regular region, characterized by smooth behavior of key quantities, 
and the critical region, where these quantities exhibit singularities or strong fluctuations. 
Due to their starkly different properties, these regions are often perceived as being weakly 
related, if at all. However, we demonstrate that these regions are intimately connected, 
a relationship that can be explicitly revealed using self-similar approximation theory.
This framework enables the prediction of observable quantities near the critical point based 
on information from the regular region and vice versa. Remarkably, the method relies solely 
on asymptotic expansions with respect to a parameter, regardless of whether the expansion 
originates in the regular or critical region. The mathematical principles of self-similar 
theory remain consistent across both cases. We illustrate this connection by extrapolating 
from the regular region to predict the existence, location, and critical indices of a 
critical point of an equation of state for a statistical system, even when no direct 
information about the critical region is available. Conversely, we explore extrapolation from 
the critical to the regular region in systems with discrete scale invariance, where 
log-periodic oscillations in observables introduce additional complexity. Our findings provide 
insights and solutions applicable to diverse phenomena, including material fracture, stock 
market crashes, and earthquake forecasting.

\end{abstract}

{\parindent=0pt
\vskip 5mm
$^a$corresponding author e-mail: yukalov@theor.jinr.ru

\vskip 2mm

 }

\newpage

\section{Introduction}

For statistical and nonlinear systems exhibiting critical phenomena, two fundamentally 
distinct regions of their characteristic parameters are generally recognized. The first 
is the region where the system displays smooth behavior of its observable quantities. 
The second is the critical region, characterized by sharp variations in certain observable 
quantities, often accompanied by oscillations and singular points. Due to these contrasting 
behaviors, these regions often appear unrelated, making it seem impossible to predict the 
behavior in one region based on knowledge of the other.

This paper argues that the regular and critical regions are deeply interconnected. When 
analyzing a specific observable quantity, let us remember that it retains the same defining 
properties across all values of the variable or parameter in question. These properties might 
be prominently exhibited in one region while remaining hidden, yet still present, in the other.

Critical phenomena are marked by scale invariance, which dictates the asymptotic behavior 
of studied quantities in the critical region 
\cite{Fisher_1,Brout_2,Stanley_3,Yukalov_4,Sornette_5,Bogolubov_6}. Typically, the behavior 
of observables in these regions is described by renormalization group equations, which leads 
to power law behavior as a specific case of continuous real-valued scaling 
\cite{Wilson_7,Hu_8,Bogolubov_9,Kadanoff_10,Ma_11,Efrati_12,Yukalov_13,Dupuis_27}. However, 
observables far outside the critical point may retain scaling properties, albeit in a hidden 
form.

Similarly, critical phenomena associated with discrete scaling involve complex critical 
exponents, whose observable signatures manifest as log-periodic oscillations superimposed 
on the leading power-law behavior \cite{Sornette_5,Sornette_14}. These oscillatory features 
may persist beyond the critical point, though they might not be readily apparent.

The possibility of log-periodic oscillations was mentioned by Harris \cite{Harris_1948}.
In a work devoted to the theory of branching processes, he suggested that iterating a 
discrete mapping might yield log-periodic oscillations. The significance of discrete scale 
invariance lies in its broad applicability across various phenomena, including phase transitions 
and temporal critical phenomena with finite-time singularities. Discrete scale invariance, 
along with its characteristic log-periodic oscillations, has been documented in numerous 
contexts, such as models defined on hierarchical lattices \cite{Derrida,Derrida_2,Derrida_3,
Bessis,Itzykson,Costin,Derrida_4}, models on fractal structures \cite{Vallejos,Knezevic,Lessa,
Bab,Padilla,Akkermans,Dunne} or on aperiodic structures with fractal spectrum
\cite{Luck,Karevski,Andrade,Carpena}, earthquake physics 
\cite{Johansen-Sammis-95,Johansen-wakita_96}, material fracture 
\cite{Anifrani-sor-95,Johansen_sor-1998,Johansen_15,Moura_16,Yukalov_17}, financial bubbles 
\cite{Sornette_bou96,Sornette_18,Johansen_19,Johansen_20,Sornette_21,Feigenbaum_2001,
sorn-book-crash-02,Feigenbaum_22,Clark_2004,Bree_2013,Fantazzini_2013,Gustavsson_2016,Ko_2018,
Chen_2018,Jhun_2018,Dai_2018,Song_2020,Shu_2024,Oguzhan_2025}, and many other systems 
\cite{Sornette_14,Yamakov,Sienkiewicz,Faillettaz,Khamzin,Bazak,Thiem,Rybczynski,Wang,Wang_2,
Bhoyar,Banerjee,Dorbath,Luck_2024}. Consequently, the derived solutions connecting regular 
and critical regions have wide-ranging applicability.

Technically, the bridge between the regular and critical regions is constructed using 
self-similar approximation theory, which has been recently reviewed in 
\cite{Yukalov_13,Yukalov_2021} (see also \cite{Yukalov_PRD,Yukalov_APNY}). This approach 
allows for the extrapolation of an observable quantity from one region to another, using only 
an asymptotic expansion available in one region. For instance, starting with an asymptotic 
expansion for small values of a variable $x \to 0$, the behavior of the observable at finite 
and even large values of $x$ (including $x \to \infty$) can be determined. This is achieved 
by extracting the self-similar properties of the expansion terms and extending this 
self-similarity to other regions.

From a mathematical standpoint, it makes no difference whether the initial expansion is 
derived in the regular or the critical region, because a suitable change of variables 
allows one to transition from small to finite or large values. What truly matters is the 
overall quality of the expansion -- namely, how many terms are included and how well those 
terms are defined.

The paper is organized as follows. In Section 2, to strengthen the completeness and 
persuasiveness of this paper, we first outline the fundamental principles of self-similar 
approximation theory and then describe the steps taken to perform the extrapolation. Section 3 
offers an explicit example that verifies the capability of self-similar approximation theory 
to bridge the regular and critical regions for the equation of state of a statistical system. 
This example illustrates that, even starting from a regular region far removed from the critical 
point and even without prior knowledge of its existence, the theory delivers accurate 
quantitative predictions for both the location of the critical point and the associated 
critical exponent. Section 4 presents the construction of a bridge extrapolating the 
process that links the critical region of a system, exhibiting discrete scale invariance, 
with its regular region. Finally, Section 5 outlines the conclusions.

\section{Self-Similar Approximation Theory}
   
The method we employ for extrapolating asymptotic expansions from one region of the 
variable to another is based on self-similar approximation theory. Naturally, there is 
no need to repeat the full mathematical foundation of the theory here, as it has been 
extensively detailed in several papers and review articles \cite{Yukalov_13, Yukalov_2021}. 
However, it is useful to briefly revisit the core concepts of the theory in order that 
the reader would understand why the self-similar approach can successfully extrapolate 
asymptotic expansions and bridge different regions with seemingly so diverse behavior. 

Suppose we are interested in a real function $f(x)$ of a real variable $x \geq 0$, which
is not known explicitly. However, we do have access to its $k$-th order expansion $f_k(x)$ 
expressed in powers of $x$ and valid in the region where $x \ra 0$. We need to extrapolate
this asymptotic expansion to the region of a finite variable $x$. This can be done by means
of self-similar approximation theory, whose main points are sketched below. We give only
the basic ideas, and the detailed mathematical description can be found in the review 
articles \cite{Yukalov_13,Yukalov_2021}.

\begin{enumerate}

\item
{\it Implantation of control parameters}. The given asymptotic series, represented by 
the truncated expansion $f_k(x)$, is typically divergent. To improve its convergence 
properties, we introduce control parameters $u$ to obtain an expanded form $F_k(x,u)$. 
These control parameters can be incorporated in various ways, including through initial 
conditions, variable transformations, or series transformations. This incorporation 
induces the convergence behavior of the series in subsequent steps.

\item
{\it Fractal transformation technique}. An efficient method of implanting control 
parameters is through the fractal transform
$$
 F_k(x,\{ n_j\} ) \; = \; \prod_{j=1}^k x^{-n_j} ( 1 + b_j x) \; ,
$$
where $n_j$ and $b_j$ are control parameters to be defined from training conditions. 

\item
{\it Determination of control functions}. To induce the convergence of the series 
$\{F_k(x,u)\}$, the control parameters, in general, have to become control functions 
$u_k(x)$, which are defined by optimization conditions, such as the minimization of
a cost functional, or training conditions so that the sequence $\{F_k(x,u_k(x))\}$ be 
convergent.

\item
{\it Construction of approximation cascade}. The transition from an expanded form 
$F_k$ to the next expanded form $F_{k+1}$ can be interpreted as the temporal motion 
within a functional space in discrete time represented by the approximation order $k$. 
The series $\{F_k(x,u_k(x))\}$ can be reformulated as a trajectory $\{y_k(f)\}$ of a 
dynamical system in discrete time, called cascade. The evolution equation of the 
approximation cascade, in the vicinity of a fixed point, is given by the self-similar 
relation $y_{k+p}(f) = y_k(y_p(f))$.

\item
{\it Embedding of cascade into flow}. In order to pass from discrete time to continuous 
time, the approximation cascade can be embedded into the approximation flow, whose 
trajectory passes through all points of the cascade trajectory and the evolution 
equation satisfies the self-similar relation $y(t+t',f) = y(t,(y(t',f))$. 

\item
{\it Integration of evolution equation}. The self-similar relation can be reformulated 
as a Lie differential equation, which can then be integrated and analyzed to define 
fixed-point solutions. These solutions serve as the effective limit $f_k^*(x)$ for the 
approximation sequence. The effective limit $f_k^*(x)$, representing a fixed point, is 
termed a self-similar approximant of order $k$.  

\end{enumerate}

Let the small-variable expansion with respect to $x > 0$ be
\be
\label{A1}
f_k(x) \; = \; \sum_{n=0}^k a_n x^n \qquad ( x \ra 0) \;   ,
\ee
where $a_0 \neq 0$. And we wish to extrapolate this expansion to finite values of 
$x$, and maybe even to large values. Self-similar approximants, derived using the 
aforementioned procedure, are found to have the form of the factor approximants 
\be
\label{A2}
f_k^*(x)  \; = \;  a_0 \prod_{j=1}^{N_k} ( 1 + A_j x)^{n_j} \; ,
\ee
in which
\begin{eqnarray}
\nonumber
N_k \; = \; \left\{ \begin{array}{rl}
k/2 \; , ~ & ~ k = 2,4, \ldots \\
(k+1)/2\; . ~ & ~ k = 3,4, \ldots 
\end{array} \right.
\end{eqnarray}
while $A_j$ and $n_j$ are control parameters defined by the training conditions
\be
\label{A3}
\lim_{x\ra 0} \;\frac{1}{n!} \;\frac{d^n}{dx^n} \; f_k^*(x) \; = \; a_n \; .
\ee

The function $f^*_k(x)$ possesses a critical point $x_c$ if there exists $m\in (0,k]$
for which 
\be
\label{A4}
 A_m \; = \; \min_j A_j \; < \; 0 \;  ~~{\rm and}~\qquad n_m \; < \; 0 \;  ,
\ee
so that $1+A_m x_c=0$, and the function $f^*_k(x)$ exhibits divergence at $x=x_c$. 
In the critical region, the function $f^*_k(x)$ has the form
\be
\label{A5}
f_k^*(x) \; \simeq \; \frac{C_k}{(x_c-x)^\al} \; ,
\ee
with the critical point and critical exponent given by
\be
\label{A6}
  x_c \; = \; \frac{1}{|\; A_m\; |} \; , \qquad 
\al \; = \; |\; n_m \; | \; ,
\ee
the coefficient $C_k$ being
\be
\label{A7}
 C_k \; = \; a_0 \prod_{j \neq m}^{N_k} ( 1 + A_j x_c)^{n_j}  \;  .
\ee
Note that, if an expansion with respect to large variables, say $z \gg 1$, is given, 
then it can be easily transformed to an expansion over small variables by means of 
the variable change $z = 1/x$.   

An important case that is often met is when the available information is minimal, 
say the expansion is 
\be
\label{A8}
 f(x) \; \simeq \; a_0 + a_1 x + a_2 x^2 \;  ,
\ee
with nonzero $a_0$ and $a_1$, while $a_2 \ra 0$. 
The optimal self-similar approximant for the given minimal information can be determined 
by applying the techniques described above to the truncated series (\ref{A8}), resulting 
in the self-similar factor approximant.
\be
\label{A9}
f^*(x) \; = \; a_0 ( 1 + A x)^n \;   ,
\ee
where $n$ is different from $0$ or $1$. From the training conditions (\ref{A3}), we have
$n = a_1/a_0 A$ and $A \ra 0$. Taking into account that
$$
 \lim_{A\ra 0} ( 1 + A x)^{1/A} \;= \; e^x \;  ,
$$
we obtain
\be
\label{A10}
 f^*(x) \; = \; a_0 \exp\left( \frac{a_1}{a_0} \; x \right) \;  .
\ee
With additional expansion terms for $f_k(x)$, more refined approximations for $f^*_k(x)$
can be achieved, with higher-order terms providing greater accuracy. However, given 
the available minimal information, the best possible result is the extrapolation given 
by approximant (\ref{A10}).

To conclude this section, let us summarize the key insight of self-similar approximation 
theory. Although perturbative series and their coefficients are typically derived for a 
specific region of a variable, they inherently contain deeper information about the full 
function they approximate, since the series are generated by that very function. The 
mathematical techniques outlined here allow us to uncover and leverage this hidden structure 
within asymptotic series. This explains intuitively why self-similar approximation methods 
can effectively extrapolate results obtained in one domain of a variable to another, broader 
domain.

For practical applications of the theory to specific problems, there is no need to revisit 
the entire underlying methodology. Instead, the essential procedure is straightforward. 
Consider an expansion (\ref{A1}) that holds in the region of small values of the variable $x$, 
which we seek to extend to finite or even large values. The solution depends on the number 
of terms available in the given expansion (\ref{A1}). In the simplest case, where only a few 
terms are known, the approximant is given by (\ref{A10}). When more terms are available, the 
appropriate extrapolating form is the self-similar factor approximant (\ref{A2}), with 
parameters $A_j$ and $n_j$ determined by the training condition (\ref{A3}). This condition 
explicitly defines all parameters $A_j$ and $n_j$. If these parameters satisfy the condition 
(\ref{A4}), then a critical point exists, leading to the approximant (\ref{A5}). This concise 
formulation encapsulates everything required for the practical implementation of the general 
theory.
   
\section{Bridging the Regular and Critical Regions for Statistical Systems}

In statistical systems, reliable expansions are typically available for their regular regions, 
far from critical points, whereas the primary interest often lies in understanding the 
critical behavior. As a typical example, let us consider the equation of state for a fluid 
of hard discs of diameter $\sigma$. The equation of state is presented by the compressibility 
factor 
\be
\label{B1}
Z \; = \; \frac{P}{\rho k_B T} \qquad
\left( \rho \equiv \frac{N}{V} \right) \;   ,
\ee
in which $P$ is pressure, $\rho = N/V$ is the density, and $T$ is the temperature. By means of 
perturbation theory, the compressibility factor can be found as an expansion in powers 
of density,
\be
\label{B2}
 Z \; = \; 1 + \frac{1}{\rho} \sum_{n=2}^\infty B_n \rho^n 
\qquad ( \rho \ra 0 ) \;  ,
\ee
where $B_n$ are the virial coefficients 
\cite{Mayer,Feynman,Hansen,Santos_1995,Clisbi_2006,Mulero_2009,Maestre_2011}. This expansion 
describes the regular region of asymptotically small density $\rho \ra 0$ located far outside 
of the critical region that is located near the closest packing density 
$\rho_c \sigma^2 = 2/ \sqrt{3} = 1.1547$. The principal question is whether it could be 
possible to predict the critical behavior of the fluid knowing only the above asymptotic 
expansion valid solely for the regular region far outside of the critical point of the 
closest packing density. 

It is convenient to use dimensionless quantities expressing the compressibility factor 
as a function of the packing fraction
\be
\label{B3}
x \; \equiv \; \frac{1}{2} \; \rho B_2 \; = \; \frac{\pi}{4} \; \rho \sgm^2 \;   .
\ee
Then the asymptotic expansion (\ref{B2}) reads as
\be
\label{B4}  
Z_k \; = \; 1 + \sum_{n=1}^k b_n x^n \qquad ( x \ra 0 ) \;   ,
\ee
where the coefficients 
$$
b_n \; = \; 2^n \; \frac{B_{n+1}}{B_2^n}
$$
take the values 
$$
b_1 = 2 \; , \qquad b_2 = 3.1280 \; , \qquad b_3 = 4.2579 \; ,  \qquad
b_4 = 5.3369 \; , \qquad b_5 = 6.3630 \; ,  
$$
$$
b_6 = 7.3519 \; , \qquad b_7 = 8.3191 \; , \qquad b_8 = 9.2721 \; , \qquad
b_9 = 10.2163 \; .  
$$

\begin{table}[ht]
\caption{Prediction for the critical point $x_c$ and
critical exponent $\al$, based on the self-similar
summation of the asymptotic expansion at $x\ra 0$ where $x$ is defined by expression 
(\ref{B3}).}
\vskip 5mm
\centering
\renewcommand{\arraystretch}{1.2}
\begin{tabular}{|c|c|c|} \hline
$k$   &  $x_c$  &  $\al$     \\ \hline
2     & 0.887   &  1.773     \\ 
3     & 0.999   &  2.124       \\ 
4     & 1.138   &  2.957     \\ 
5     & 1.092   &  2.591   \\
6     & 0.998   &  1.831     \\ 
7     & 0.993   &  1.783         \\
8     & 0.994   &  1.794    \\
9     & 0.994   &  1.790  \\ \hline
\end{tabular}
\end{table}

Summing expansion (\ref{B4}) by means of the self-similar approximation theory using the
self-similar factor approximants    
\cite{Yukalov_13,Yukalov_2021,Yukalov_2003,Gluzman_2003,Yukalov_2022}, we find the 
equation of state
\be
\label{B5}
 Z_k^* \; = \; \frac{C_k}{(x_c-x)^\al} \;  ,
\ee
in which the coefficient $C_k$ converges to $C = 1.54$ for large $k$. The critical 
normalized density $x_c$ and the critical exponent $\alpha$ for different approximation 
orders $k$ are given in Table 1. The critical point converges to $x_c = 0.99$ and the 
critical exponent, to $\alpha = 1.79$, which is in perfect agreement with the values 
obtained in numerical calculations \cite{Clisbi_2006}.      

Thus, the self-similar approximation theory not only enables us to connect the critical 
region to the distant regular region, but it also provides precise quantitative 
predictions for the location of the critical point and the value of the critical exponent.

\section{Bridging the Critical and Regular Regions for Systems with Discrete Scale 
Invariance}

For systems with discrete scale invariance, the situation is the reverse of what was described 
in the previous section. There are quite numerous studies for the critical region and 
practically no information on the regular region. It is therefore necessary to start with 
the critical region and try to realize an extrapolation towards the regular region.  

In the critical region, accomplishing the decimation procedure for lattice spin systems 
\cite{Hu_8,Kadanoff_10,Efrati_12}, it has been shown \cite{Niemeijer_1976,Derrida}
that, in the vicinity of the phase transition point, the existence of discrete scale
invariance leads to a specific renormalization group equation for the system free energy.
The latter can be considered as a function of temperature, or of an effective coupling
$q=\exp(J/T)$, where $J$ is a strength of spin interactions and $T$ is the temperature.
Using the dimensionless relative coupling $x = 1 - q/q_c$, where $q_c=\exp(J/T_c)$, 
with $T_c$ being a critical temperature, consider the decimation of scale $\lambda > 1$ and 
the related renormalization of $x$
\begin{equation}
 \hat R (x)~ = ~ \lbd x \; , \qquad \hat R^n(x) ~ = ~\lbd^n x \;  .
\end{equation}
The corresponding free energy per lattice site has been shown \cite{Niemeijer_1976,Derrida}
to satisfy the equation
\be
\label{1}
 f(x) ~ \simeq ~ h_0(x) + \frac{1}{\mu} \; f[\; \hat R(x) \; ] \;  ,
\ee
related to the discrete scale invariance in the vicinity of the critical point, where $x$ 
is close to zero and $\mu > 1$ is the renormalization scale for the free energy. Equation 
(\ref{1}) is approximately valid only near a critical point, as each step of the decimation 
procedure typically involves the multiplication of effective couplings. The distinction 
between relevant and irrelevant couplings, which reduces their number, can generally be made 
only in the vicinity of a critical point
\cite{Wilson_7,Hu_8,Bogolubov_9,Kadanoff_10,Ma_11,Efrati_12,Yukalov_13,Dupuis_27}.

By iterating the above equation, one obtains a formal solution
\be
\label{2}
 f(x) ~ \simeq ~
\sum_{n=0}^\infty \frac{1}{\mu^n} \; h_0[\; \hat R^n(x) \; ] \; .
\ee
The existence of discrete scale invariance is related to the occurrence of periodic
or chaotic maps in the renormalization procedure \cite{Derrida}.

It has been shown \cite{Gluzman_2002} that the free energy satisfying the
renormalization group equations (\ref{1}) and (\ref{2}), connected with discrete scale
invariance, can be represented as
\be
\label{3}
f(x) ~ \simeq ~  h_0(x) + x^\al
\sum_{n=0}^\infty A_n x^{i n\om} \;   ,
\ee
where $\al=\ln\mu/ \ln \lbd$ and $\om= 2\pi/\ln\lbd$. It is easy to notice
that using the identity
$$
{\rm Re} \sum_{n=0}^\infty A_n x^{i n\om} ~ = ~
\sum_{n=-\infty}^\infty B_n x^{i n\om} \;  ,
$$
in which $B_0 = A_0$ and $B_n = B_{-n}= A_n/2$ for $n \geq 1$, the free energy (\ref{3}) can 
be rewritten in the form
\be
\label{4}
f(x) ~ \simeq ~  h_0(x) + x^\al
\sum_{n=-\infty}^\infty B_n x^{i n\om} \;  .
\ee
The appearance of complex powers leads to the arising log-periodic oscillations, since
$$
x^{i \om} ~ = ~ \cos(\om \ln x) + i \sin(\om \ln x ) \; .
$$

In order to get complex powers, it is possible to work from the beginning in the complex
plane, keeping in mind that the observable quantity is real. Assuming that, in the 
complex plane  in the vicinity of a critical point, the observable satisfies the 
renormalization group equation
\be
\label{6}
 \frac{d F(x)}{d\ln x} ~ \simeq  ~ G[\; F(x) \; ] \; ,
\ee
in which $G[F(x)]$ is the Gell-Man-Low function
\cite{Wilson_7,Bogolubov_9,Kadanoff_10,Yukalov_13,Gell_23,Bogolubov_24}, one can look
for a solution by expanding the Gell-Man-Low function in powers of $F(x)$, as has been
done in Refs.~\cite{JohSoranti99,JohSorJan00,SorZhou0002deep,JohSorfts50,ZhouSoranti03}.

In all cases, one obtains the behavior of observable quantities in the vicinity of a
critical point, where discrete scale invariance is present. Here, our aim is to 
extrapolate the critical behavior to the regular region far outside the critical vicinity.
Since there exist numerous phenomena exhibiting discrete scale invariance, we shall 
consider the behavior of their general typical representative denoted by a dimensionless
function $f(x)$. We consider this characteristic quantity $f(x)$ as a function of a 
dimensionless variable $x$. The quantity of interest can be a thermodynamic characteristic, 
an asset market price, a financial index, the intensity of acoustic emissions from a 
loaded material undergoing damage, or its amplitude of deformation, an electric current 
or some other potential precursors of impending earthquakes. Without loss of generality, 
the function $f(x)$ can be defined to be non-negative.

In the vicinity of a critical point $x_c$, the studied quantity experiences a sharp
increase. Depending on the considered particular case, the dimensionless variable
$x$ can represent relative pressure, relative elongation, relative temperature, relative
effective coupling, relative time, or other relative system parameters,
\be
\label{7}
 x ~ = ~ \left\{ 1 - \; \frac{p}{p_c} \; , ~ 1 - \; \frac{l}{l_c} \; , ~
1 - \; \frac{T}{T_c} \; , ~ 1 -\; \frac{q}{q_c} \; , ~ 1 - \; \frac{t}{t_c} \right\} \; .
\ee
In that choice of variable, the initial point equals one, while the critical
point corresponds to zero,
\be
\label{8}
x_0 ~ = ~ 1 \; , \qquad x_c = 0 \;   .
\ee
This choice is convenient, because then the critical region is characterized by
$x \ll 1$, so that in this region it is possible to use expansions in powers of $x$.
Thus the introduced relative variable takes values in the interval
\be
\label{9}
 0 ~ = ~ x_c \; \leq \;  x \; \leq \;  x_0 = 1 \; ,
\ee
while the normalised physical variable, varies in the interval
\be
\label{10}
 0 ~ = ~ 1 - x_0 \; \leq \; 1 - x \; \leq \; 1 - x_c = 1 \;   .
\ee

An observable quantity $I(x)$ can always be normalized, say defining $f(x)=I(x)/I(0)$,
so that the dimensionless form for the function of interest is positive and satisfies
the normalization
\be
\label{11}
 f(0) ~ = ~ 1 \; ,  \qquad f(x) ~ \geq ~ 0 \; .
\ee
Our aim is to study the general situation where, in the critical region in the vicinity
of a critical point, there exists discrete scaling. Then, in the critical region, the
studied function $f(x)$ can be represented as
\be
\label{12}
f(x) ~ \simeq ~ 1 + g(x) \qquad ( x \ra 0 ) \; ,
\ee
where $g(x)$ is a function that in the critical region enjoys discrete scale invariance.
When using the scale invariance relation, representation (\ref{12}) becomes equivalent 
to expression (\ref{1}). In view of the normalization condition $f(0) = 1$, we have the
boundary condition
\be
\label{13}
g(0) ~ = ~ 0 \; .
\ee

In order to find the general form of the considered function $f(x)$ in the whole
interval $x \in [0,1]$ from the knowledge of the existence in the critical region
$x\ra 0$ of discrete scaling, it is necessary to resort to the self-similar 
extrapolation procedure.

The function $g(x)$ is assumed to satisfy a general type of scale invariance yielding
complex exponents \cite{Sornette_5,Sornette_14}. At the same time, there is no need to
extend the consideration to the complex plane, since complex exponents naturally appear
after employing Fourier transformation.

The property of discrete scale invariance, arising in the vicinity of a critical point,
reads as
\be
\label{14}
 g(\lbd x)  ~ = ~ \mu g(x) \;  .
\ee
Here the scaling parameters $\lambda>1$ and $\mu>1$ are assumed to be fixed. This is
contrary to continuous scaling where these parameters are arbitrary. Repeated use of
the scaling relation yields
\be
\label{15}
  g(\lbd_n x)  ~ = ~ \mu_n g(x) \;  ,
\ee
where the scaling parameters $\lbd_n=\lbd^n$ and $\mu_n = \mu^n$ compose a discrete
spectrum with $n = 0, \pm 1,\pm 2,\ldots$.

It is easy to check that the general solution of relation (\ref{14}) has the form
\be
\label{16}
 g(x)  ~ = ~ x^\al y(x) \;   ,
\ee
in which, independently of $n$,
\be
\label{17}
\al  ~ \equiv ~ \frac{\ln\mu}{\ln\lbd}
\ee
and the function $y(x)$ can be expressed as
\be
\label{18}
y(x)  ~ = ~ F \left( \frac{\ln x}{\ln \lbd} \right) \;   ,
\ee
with $F(z)$ being a periodic function of period one,
\be
\label{19}
F(z+1)  ~ = ~ F(z) \qquad
\left( z \equiv \frac{\ln x}{\ln\lbd} \right) \;   .
\ee

A periodic function can be expanded in the Fourier series
\be
\label{20}
 F(z)  ~ = ~ \sum_{n=-\infty}^\infty B_n e^{i 2\pi n z} \;  ,
\ee
which, using the notation of $z$ in (\ref{19}), can be rewritten in the form
\be
\label{21}
 y(x)  ~ = ~ \sum_{n=-\infty}^\infty B_n e^{i n \omega \ln x} \;  ,
\ee
with the log-frequency
\be
\label{22}
 \om  ~ = ~ \frac{2\pi}{\ln\lbd} \;  .
\ee
The log-frequency $\om$ encodes the existence of the preferred discrete scaling ratio 
$\lambda$, such that $\ln \lambda$ is the period of the oscillations in the $\ln x$ variable. 
Thus, $\lambda$ is not a frequency (the inverse of time) in the usual sense but the preferred 
scaling factor. Keeping in mind a real function $F(z)$, hence real $y(x)$, gives the equality 
$B^*_n = B_{-n}$.

Thus for the sought quantity (\ref{12}) in the critical region, we have
\be
\label{23}
f(x)  ~ \simeq  ~  1 + x^\al y(x) \qquad ( x \ra 0 ) \; .
\ee
This expression is valid only in the vicinity of a critical point, where $x \ra 0$.
Therefore it is possible to consider this expression as providing the first terms of an 
expansion
\be
\label{24}
f(x)  ~ = ~ \sum_{m=0}^\infty a_m x^{m\al} y^m(x) \;   ,
\ee
where $a_0 = a_1 = 1$.

A representation in the form of infinite series  (\ref{24})  is not merely inconvenient
for practical use, but it has several principal deficiencies. Generally, such series
are asymptotic and have meaning only in the limit $x \ra 0$. For a finite value of $x$,
they may not converge and even can become negative, breaking the condition of
semi-positiveness of $f(x)$. Dealing with such a series requires to define its effective
sum. This can be done by resorting to self-similar approximation theory delineated in
Sec. 2. In the case, when just a few terms of an expansion is available and it is 
necessary to guarantee semi-positiveness of the sought function, the method leads to 
the approximant (\ref{A10}). This form guarantees the best self-similar extrapolation 
under the minimally available information. 

Note that, when more information is available, and a number of terms in an expansion 
is sufficiently large, it is possible to extend the form (\ref{A10}) by repeated
iteration leading to continued exponentials \cite{Euler_1777,Barrow,Bell,Knoebel_1981,
Rippon,Bromer,Bender_1996,Anderson,Devaney,Hooshmand,Marshall,Paulsen}, 
which finds its justification \cite{Yukalov_25} in self-similar approximation theory 
\cite{Yukalov_13,Yukalov_2021}. This type of exponential summation has been successfully 
used in several problems 
\cite{Yukalov_13,Moura_16,Yukalov_17,Bender_1996,Yukalov_25,Yukalov_2002,Gluzman_26,
Abhignan_2021,Abhignan_San_2021,Abhignan_2023}. 

Employing the self-similar approximant (\ref{A10}), the effective sum (\ref{24}) can 
be represented by the expression
\be
\label{25}
f(x)  ~ = ~ \exp \{ x^\al y(x) \} \;  .
\ee
As is mentioned above, the used approximation is based on the self-similar approximation 
theory that enjoys a sound mathematical foundation and has been demonstrated to provide 
accurate approximations for a great number of examples. 

The coefficients $B_n$ in expansion (\ref{20}) are proportional to $\mu^{-n} < 1$, which
allows us to treat $B_n$ as decreasing with $n$ (see \cite{Derrida,Gluzman_2002}).
Keeping the first three terms, we have
\be
\label{26}
 F(z)  ~ = ~ B_0 + B_1 e^{i 2\pi z} + B_1^* e^{-i 2\pi z} \;  ,
\ee
which yields
\be
\label{27}
y(x) = a + b \cos(\om \ln x) - c \sin(\om \ln x) \;   ,
\ee
where
$$
 a  ~ = ~ B_0 \; , \qquad b  ~ = ~ 2 {\rm Re} B_1 \; , \qquad
c  ~ = ~ 2 {\rm Im} B_1 \;  .
$$
Equivalently, this can be represented as
\be
\label{28}
  y(x)  ~ = ~ a + b \cos(\om \ln x + \vp) \; ,
\ee
with
$$
\vp  ~ = ~ \arctan \frac{c}{b}  ~ = ~
\arctan \frac{{\rm Im} B_1}{{\rm Re} B_1} \;   .
$$

In this way, the extrapolation of the studied observable quantity from the critical
region, where $x \ra 0$, to arbitrary values of $x$ reads as
\be
\label{29}
f(x)  ~ = ~ \exp\{ x^\al [\; a + b\cos ( \om \ln x + \vp ) \; ] \} \; .
\ee

Note that setting $b \ra 0$ corresponds to the case of continuous scale invariance,
so that function (\ref{29}) reduces to
\be
\label{30}
 f_0(x)  ~ = ~ \exp( a x^\al) \;  .
\ee

The extrapolated function (\ref{29}) represents the quantity of interest for
all $x$ in the domain $[0, 1]$. In terms of physical behavior, we have to consider
the variable $1-x$ for studying the behavior of the function $f(x)$ from the initial
point $1-x_0 = 0$ to the critical point $1-x_c = 1$.

In order to analyse the influence of discrete scale invariance, it is useful to compare
the full function $f(x)$ with the situation when only continuous
scale invariance exists, and discrete scale invariance is absent. This corresponds to
function (\ref{30}). In the vicinity of the critical point, the continuous scale
invariance leads to the power law in the exponential.

The typical behavior of $f(x)$, as compared to $f_0(x)$, as functions of $1-x$, in
the whole range $[0,1]$ is demonstrated in Figs. 1 to 5. The parameters are taken
close to the values describing the behaviour before material ruptures
\cite{Johansen_15,Moura_16,Yukalov_17} and crashes in financial markets
\cite{Sornette_18,Johansen_19,Johansen_20,Sornette_21}. The role of the parameters
entering $f(x)$ is emphasized.

Figure 1 illustrates that the parameter $\omega$ controls the number of oscillations, 
the larger $\omega$, the larger the number of oscillations. The phase shift $\varphi$, 
as is shown in Fig. 2, moves the location of the oscillations, but does not change 
the overall picture qualitatively. Increasing the parameter $\alpha$ smooths the 
log-periodic oscillations, and makes $f(x)$ larger, as is clear from Fig. 3.
Decreasing parameter $a$ diminishes the oscillation amplitude, as in Fig. 4. The
absolute value of $b$ influences the oscillation amplitude, making it larger by
increasing $|b|$, as is evident from Fig. 5.

The function $F(z)$, according to its representation (\ref{20}), contains many
modes with different frequencies. The reduction to the simple form (\ref{26}) is valid
only when $1/\mu^n \ll 1$ for $n>1$ so that the coefficients $B_n$ quickly decrease with 
$n$ and only the lowest mode is the most important.

In order to understand the influence of higher modes, it is necessary to study the
behaviour of quantity (\ref{25}) with the general form of the function
\be
\label{31}
 F(z)  ~ = ~ B_0 +
\sum_{n=1}^\infty \left( B_n e^{i 2\pi n z} + B_n^* e^{-i 2\pi n z} \right) \;  ,
\ee
which leads to
\be
\label{32}
 y(x)  ~ = ~ a + \sum_{n=1}^\infty b_n \cos ( n \om \ln x + \vp_n ) \;  ,
\ee
with $b_n = b/\mu^{n-1}$.

Note that in Ref. \cite{Gluzman_2002} it is shown that two cases can happen, one is 
when the coefficients $b_n$ decrease exponentially as a function of $n$, so that only 
a few lower modes survive, and the other case, when the coefficients $b_n$ decrease 
slower, so that many modes are to be taken into account. These two cases correspond 
to very weak and strong log-periodicity, respectively. The form $b_n = b/\mu^{n-1}$, 
depending on the value of $\mu$, can characterize these two cases. For $\mu > 1$, the 
coefficients decrease exponentially and one can limit the consideration to a few first 
modes, while for $\mu$ close to one a number of modes needs to be taken into account, 
as is shown below.   

Figure 6 shows the typical behaviour of $f(x)$, when the number $N$ of modes increases
until the function behaviour ceases to change. The phases of all modes are assumed
to be equal, $\varphi_n = \varphi$. First, increasing the number of modes makes
the oscillations more complex. When this number becomes close to $N = 25$, 
a type of mode condensation occurs, characterized by shapes resembling deep troughs. 
After $N = 70$, the behaviour of $f(x)$ practically does not change. Summarizing the 
information in Fig. 6, we see that: The log-periodic oscillations become more and more 
complex with increasing the number of modes $N$. A mode condensation effect starts as
the number of modes reaches $N=25$. Beyond $N=70$, the function stabilizes and does not 
change significantly, indicating that higher modes do not essentially alter the function.  

Varying the phase $\varphi$ can result in the appearance of high peaks before the abrupt 
falls, as in Fig. 7. The phase shift affects the alignment of oscillations. Overall, phase
shift does not change much the log-periodic structure of oscillations merely relocating 
the oscillation peaks and slightly smoothing the latter. These variations may be useful 
in tuning models for specific applications, e.g., for the prediction of financial crashes
or material rupture.

Decreasing the absolute value of negative $a$ makes the peaks smoother, as in Fig. 8, as 
compared with these in Fig. 7. Smaller negative $a$ suppresses the peaks and troughs, 
compared to larger negative values. Again, as in Fig. 7, the function stabilizes at $N=70$,
which confirms that $N=70$ is sufficient to capture full oscillation complexity. 

The exponent $\alpha$ controls how quickly log-periodic oscillations grow. Smaller 
$\alpha = 0.3$ in Fig.9, as compared with $\alpha = 1$ in Fig. 8, leads to smaller 
amplitudes of the oscillations and slower growth, as is shown in Fig. 9. The function 
is stable with respect to the increasing number of modes $N$ after $N=70$ confirming 
the robustness of the self-similar approximation. 

The parameter $\omega$ controls the number of log-periodic oscillations. Figure 10 is 
plotted for larger $\omega$, as compared to Fig. 9, which results in more frequent 
oscillations. Influencing how densely spaced the oscillations are impacts how observable 
log-periodic patterns appear in empirical data. 
 
Figures 11 and 12 demonstrate the influence of randomized phase effects on log-periodic 
oscillations. Random phases are drawn from normal distributions with different standard
deviations. When phase values are randomized, oscillations become more chaotic. With 
increasing number of modes $N$ oscillations become increasingly irregular. For high $N$,
irregularity dominates, resembling rather a chaotic dynamics than a structured log-periodic 
pattern. Thus randomness significantly affects the structure of fluctuations, with higher 
standard deviations leading to less predictable and more chaotic behavior. This picture is
relevant for modeling natural phenomena where oscillations are disrupted by random noise.  

Concluding, these figures explore how different parameters of the model influence 
log-periodic oscillations. Highlighting the key points it is possible to summarize the 
following. Increasing the number of modes $N$ enhances oscillation complexity. Beyond the
threshold $N=70$, further mode contributions are getting practically negligible. Varying
the phase shifts $\varphi$ moves the oscillations, but does not change their fundamental 
nature. The amplitude parameter $a$ determines how pronounced the oscillations appear.
The exponent $\alpha$ governs the overall growth rate of oscillations. Log-frequency 
$\omega$ directly influences how many oscillations appear over a given range. Phase 
randomness disturbs and even can destroy the standard picture of log-periodic oscillations,
mimicking real-world systems where the main trends are usually superimposed by random
noise.

\section{Conclusion}

We have considered the relation between the regular region far outside of a critical 
point and the critical region in the vicinity of the critical point. Despite rather
different behavior of observable quantities in these regions, they are not completely
disconnected, but there is a bridge between them. We put forward the idea that the
connection between the regular and critical regions can be recovered on the basis of
self-similarity in the variation of their asymptotic forms in these regions. It is 
possible, starting with an asymptotic expansion in one of the regions, to explicitly 
extrapolate this expansion to another region. This can be accomplished employing the 
self-similar approximation theory, thus creating a self-similar bridge between the 
regions.

From a mathematical perspective, it is important to note that it makes no difference 
whether the starting point is the regular region or the critical region. This is because 
what we need is the existence of an expansion in powers of some parameter. The theory allows 
for the extrapolation of any expansion, irrespectively of the nature of the parameter. 
Of course, it is reasonable to start from that region where there is a better defined 
expansion.

As an example of self-similar extrapolation from the regular to critical region, we 
have considered a statistical system, represented by a fluid of hard disks. We have 
shown that, on the basis of an expansion in the regular region, being far outside the 
critical region and even having no information on its existence, it is straightforward 
to extrapolate the equation of state to the critical point and, moreover, to predict 
the location of the critical point and to find the related critical exponent.

Another example we studied involves systems with discrete scale invariance. In this case, 
while more information is available on the system's critical behavior, the regular region 
remains largely unexplored. This presents the challenge of extrapolating from the critical 
region to areas far beyond it. This issue is common across various critical phenomena 
exhibiting discrete scale invariance, such as earthquakes, material fractures, stock market 
crashes, and other ``rupture'' events.

A key conclusion to highlight is that log-periodic oscillations observed near a critical 
point do not emerge spontaneously. Rather, they have existed well outside the critical region, 
though typically smaller in amplitude, less frequent, and smoother. This indicates that it 
is possible to identify and study log-periodic oscillations even at significant distances from 
the critical point. Furthermore, for certain parameter settings, these oscillations outside 
the critical region can exhibit quite large amplitudes. Varying the parameters can result 
in vastly different behaviors, providing models for a wide range of processes, including the 
aforementioned earthquakes, material fractures, and stock market crashes.

The examples analyzed reinforce the concept that the regular and critical regions are deeply 
interconnected. This relationship can be uncovered using self-similar approximation theory, 
which enables the construction of a self-similar bridge linking the regular and critical 
regions. This conclusion is broad in scope and applies to systems of various types and natures.

\section*{Statements and Declarations}

{\parindent=0pt

{\bf Funding/Competing Interests}: D.S. was partially supported by the National Natural 
Science Foundation of China (Grant No. T2350710802 and No. U2039202), Shenzhen Science and 
Technology Innovation Commission Project (Grants No. GJHZ20210705141805017 and No. K23405006), 
and the Center for Computational Science and Engineering at Southern University of Science 
and Technology.

\vskip 2mm
{\bf Financial interests}: The authors have no relevant financial or non-financial 
interests to disclose.

\vskip 2mm
{\bf Conflicts of interests}:
The authors declare no conflicts of interest.

\vskip 2mm
{\bf Author contributions}: All authors contributed to the study conception and design.
Material preparation, data collection and analysis were performed by V.I. Yukalov,
E.P. Yukalova, and D. Sornette. The first draft of the manuscript was written by
V.I. Yukalov and all authors commented on previous versions of the manuscript. All 
authors read and approved the final manuscript.

\vskip 2mm
{\bf Data Availability Statement}: No Data associated in the manuscript.

}

\newpage

\begin{figure}[ht]
\centerline{
\hbox{
\includegraphics[width=7.5cm]{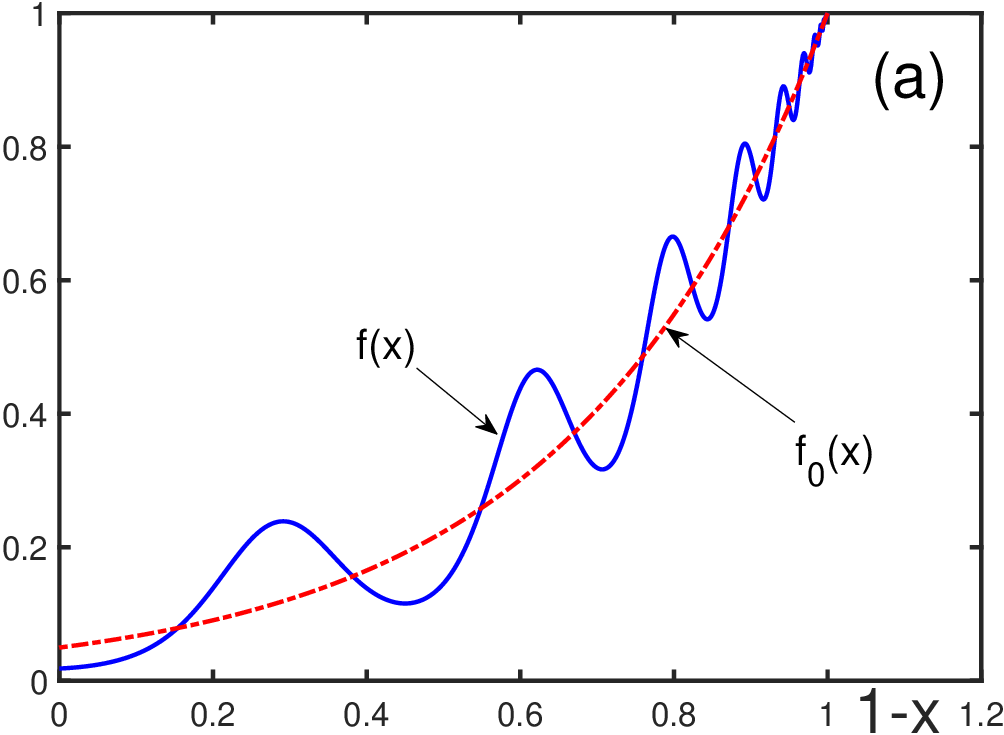} \hspace{1cm}
\includegraphics[width=7.5cm]{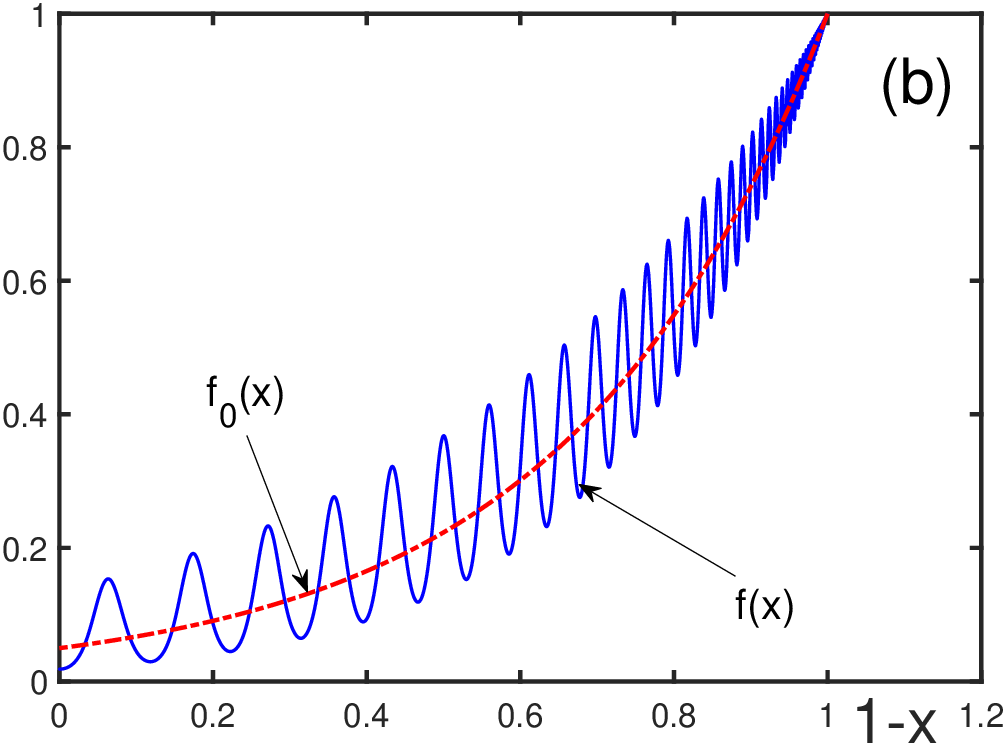} } }
\caption{\small
Functions $f(x)$ (solid line) and $f_0(x)$ (dashed-dotted line) for the fixed
parameters $\al=1$, $a=-3$, $b=-1$, $\vp=0.1$, and
$\om$ being varied:
(a) $\om=10$;
(b) $\om=50$;
}
\label{fig:Fig.1}
\end{figure}

\vskip 5cm

\begin{figure}[ht]
\centerline{
\includegraphics[width=10cm]{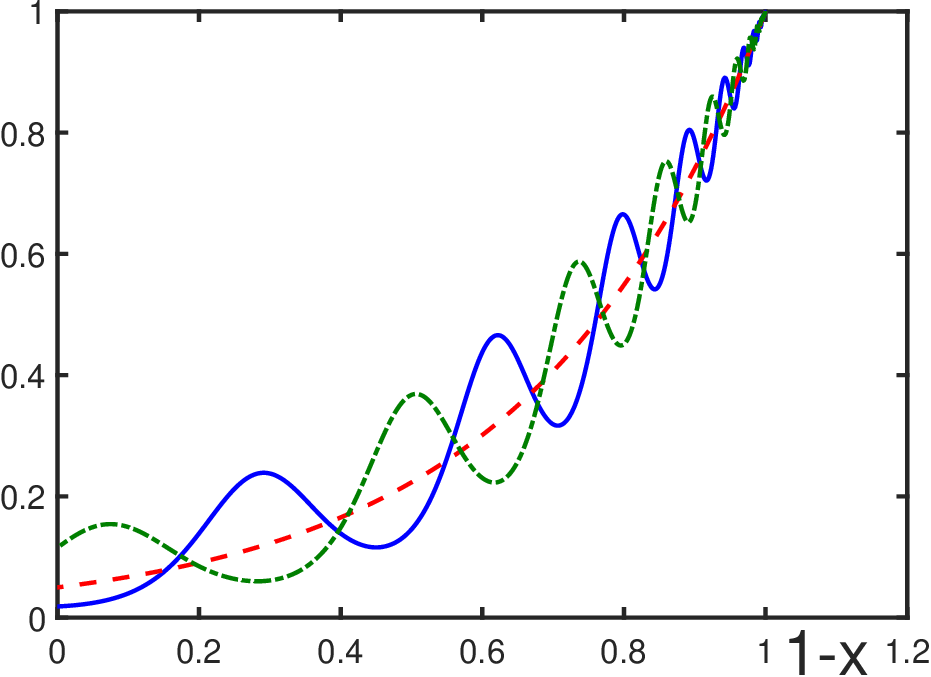}  }
\caption{\small
Functions $f(x)$ (solid blue line) for $\varphi=0.1$, $f(x)$ (dashed-dotted green line)
for $\varphi=10$, and function $f_0(x)$ (dashed red line) are presented for the
fixed parameters  $\al=1$, $\bt=-1$, $a=-3$, $b=-1$, and $\om=10$. Note that
$f_0(x)$ does not depend on $\varphi$.
}
\label{fig:Fig.2}
\end{figure}

\vskip 5cm

\begin{figure}[ht]
\centerline{\hbox{
\includegraphics[width=7.5cm]{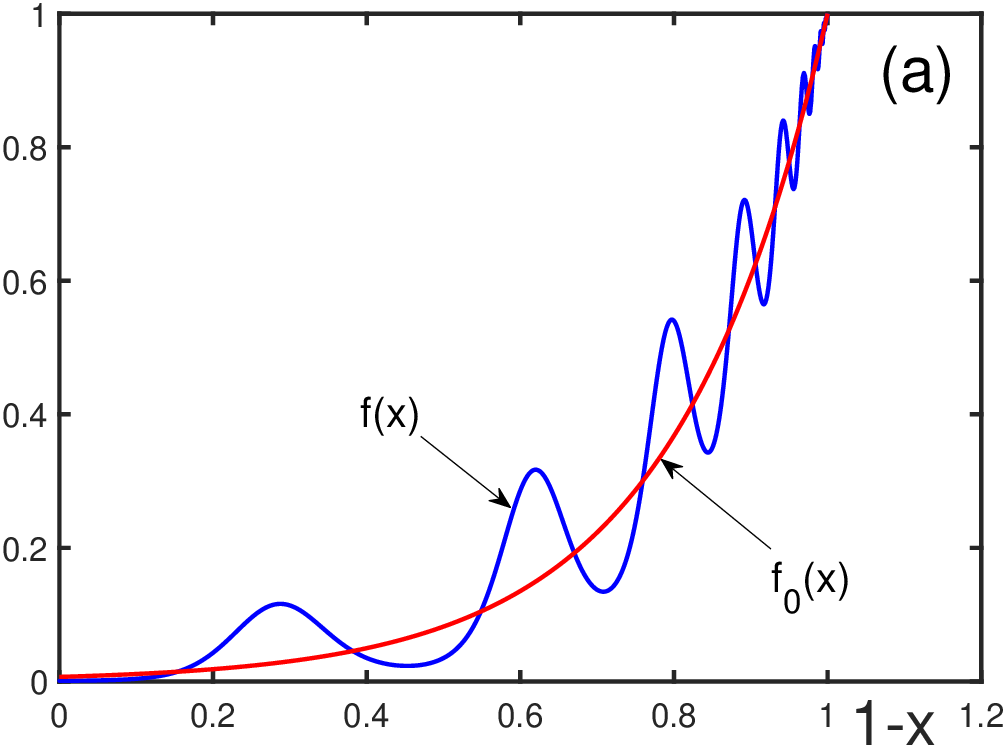} \hspace{1cm}
\includegraphics[width=7.5cm]{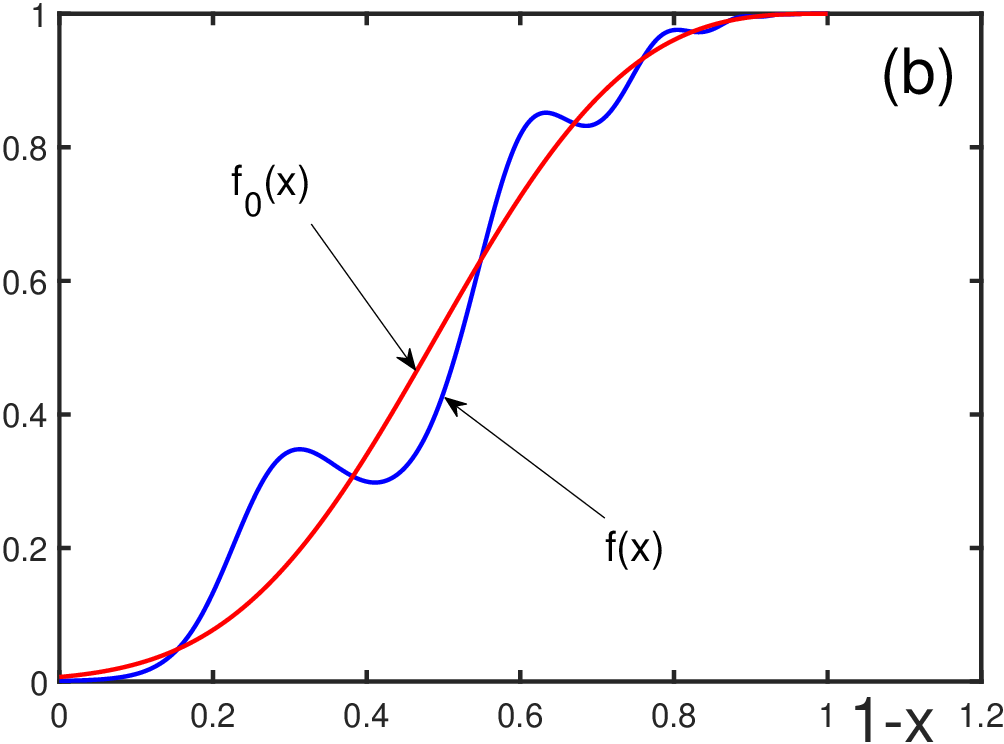} }   }
\caption{\small
Functions $f(x)$ (oscillating solid blue line) and $f_0(x)$ (monotonic solid red line) 
for the fixed parameters $a=-5$, $b=-2$, $\om=10$, $\varphi=0.1$, and $\al$ being varied:
(a) $\al=1$;
(b) $\al=3$;  
}
\label{fig:Fig.3}
\end{figure}

\vskip 5cm

\begin{figure}[ht]
\centerline{\hbox{
\includegraphics[width=7.5cm]{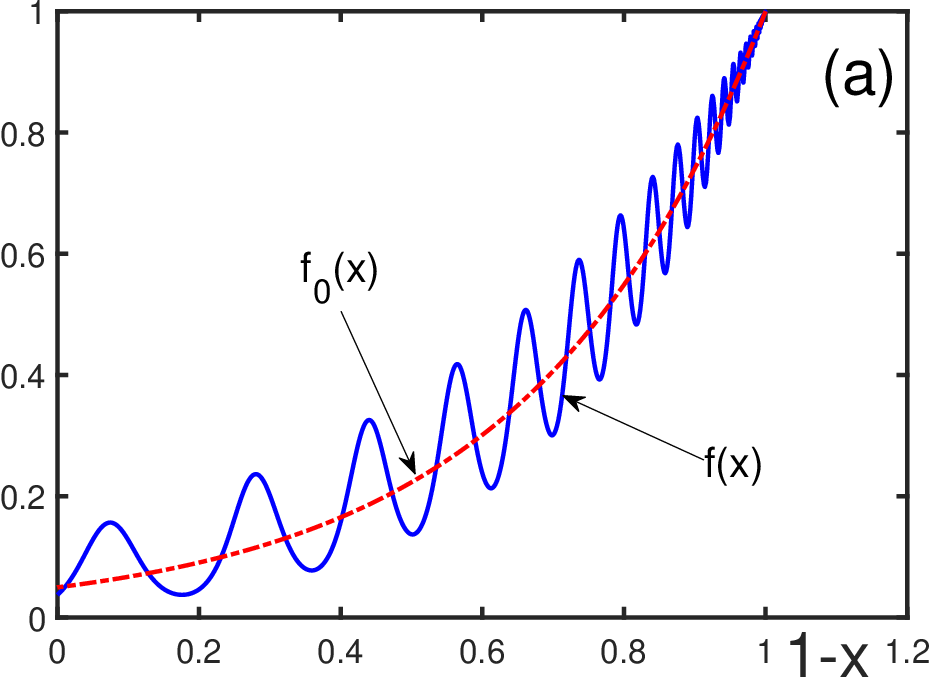} \hspace{1cm}
\includegraphics[width=7.5cm]{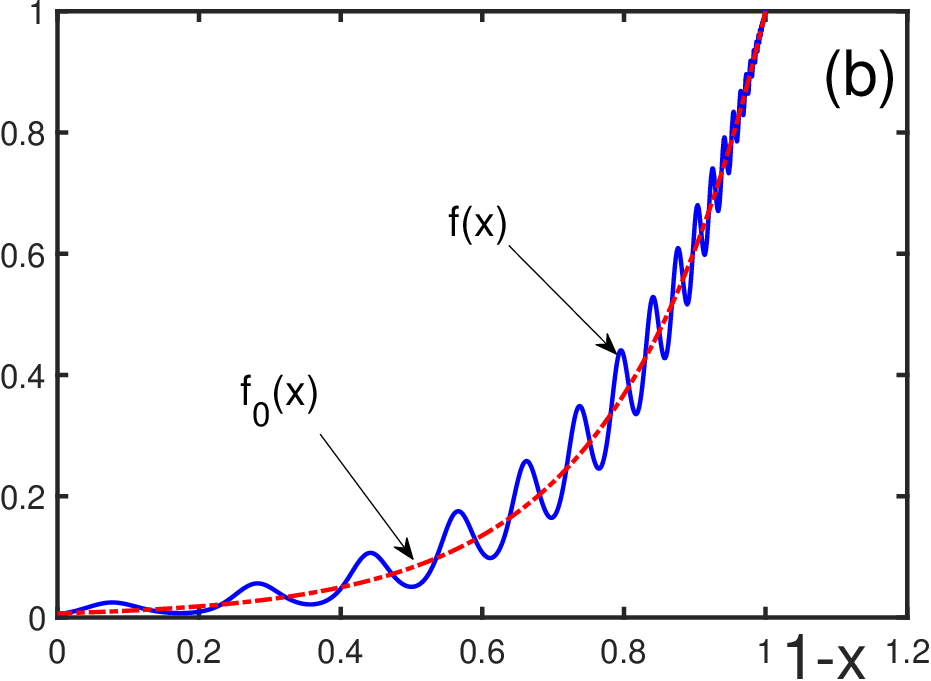} }   }
\caption{\small
Functions $f(x)$ (solid blue line) and $f_0(x)$ (dashed-dotted red line) for the fixed
parameters $\al=1$, $b=-1$, $\om=25$, $\varphi=5$, and $a$ being varied:
(a) $a=-3$;
(b) $a=-5$;
}
\label{fig:Fig.4}
\end{figure}

\vskip 5cm

\begin{figure}[ht]
\centerline{\hbox{
\includegraphics[width=7.5cm]{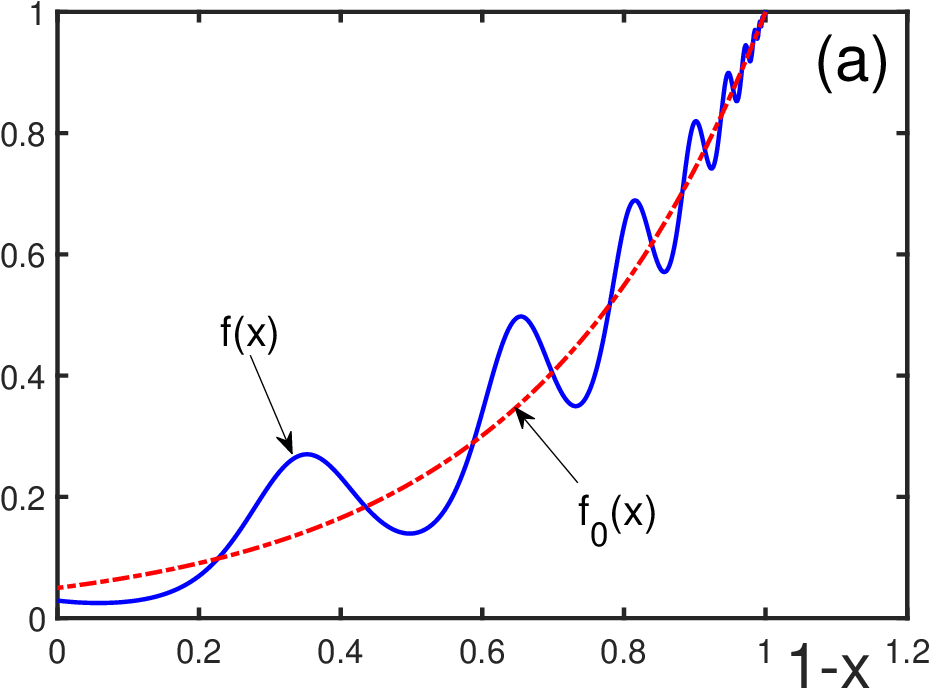} \hspace{1cm}
\includegraphics[width=7.5cm]{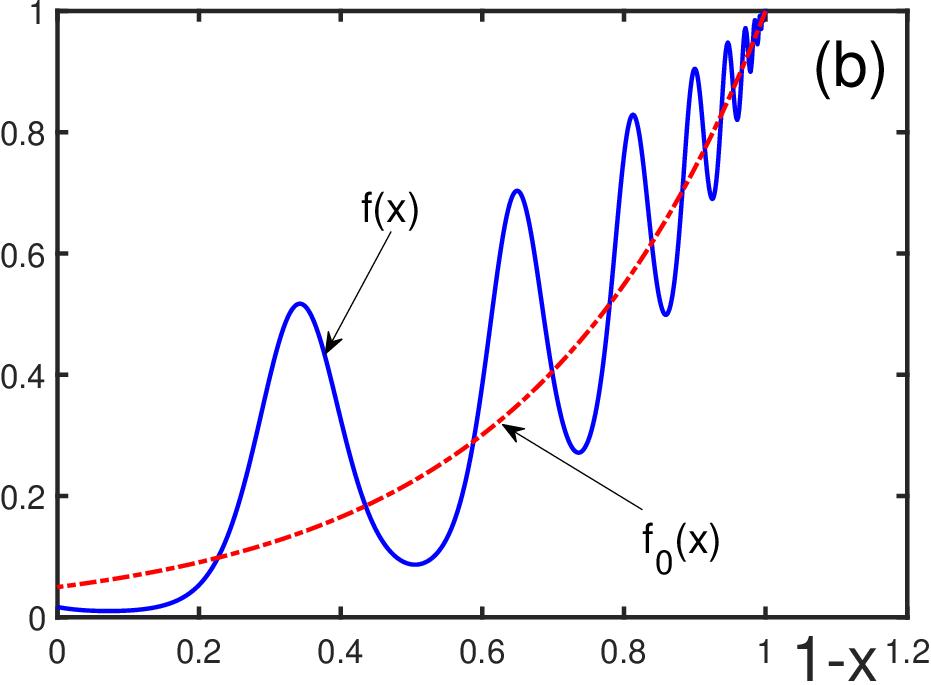} }   }
\caption{\small
Functions $f(x)$ (solid blue line) and $f_0(x)$ (dashed-dotted red line) for the fixed
parameters $\al=1$, $a=-3$, $\om=10$, $\varphi=1$, and $b$ being varied:
(a) $b=-1$;
(b) $b=-2$;
}
\label{fig:Fig.5}
\end{figure}

\vskip 5cm

\begin{figure}[ht]
\centerline{
\hbox{
\includegraphics[width=7.5cm]{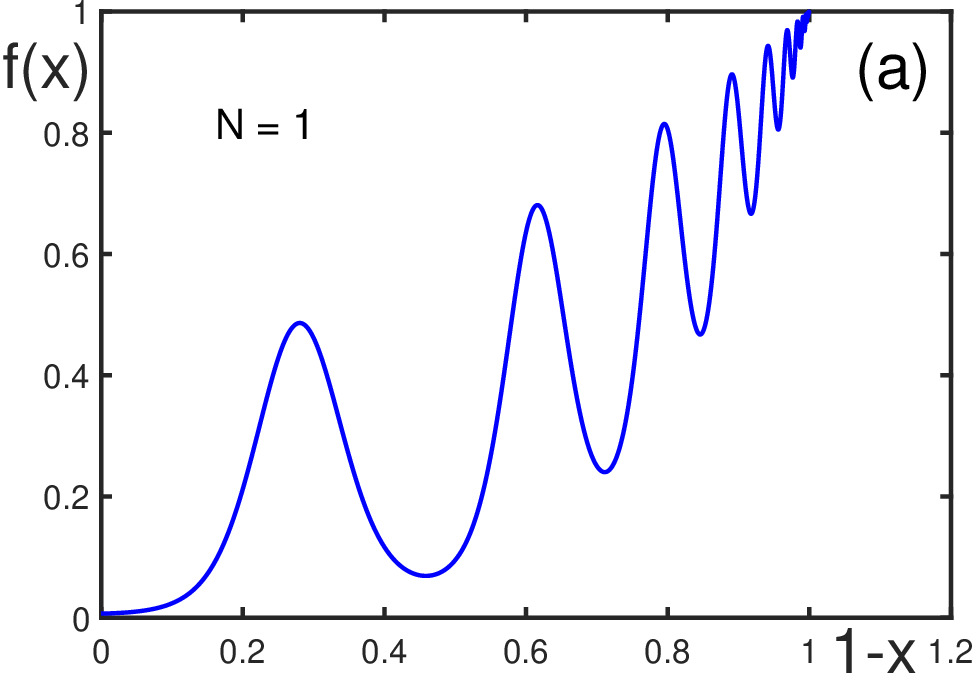} \hspace{1cm}
\includegraphics[width=7.5cm]{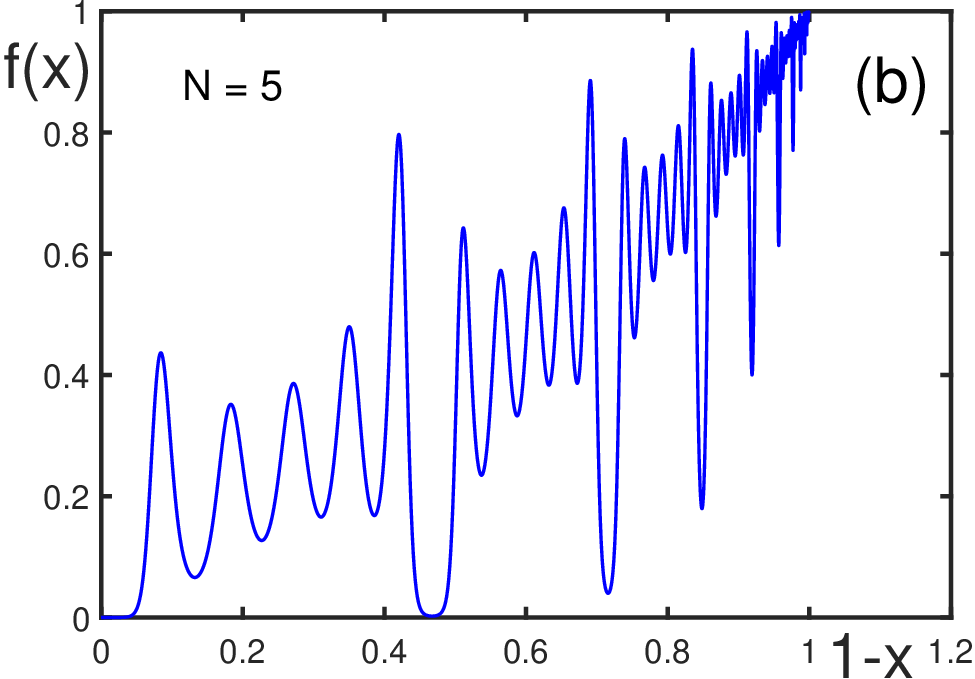} } }
\vskip 1cm
\centerline{
\hbox{
\includegraphics[width=7.5cm]{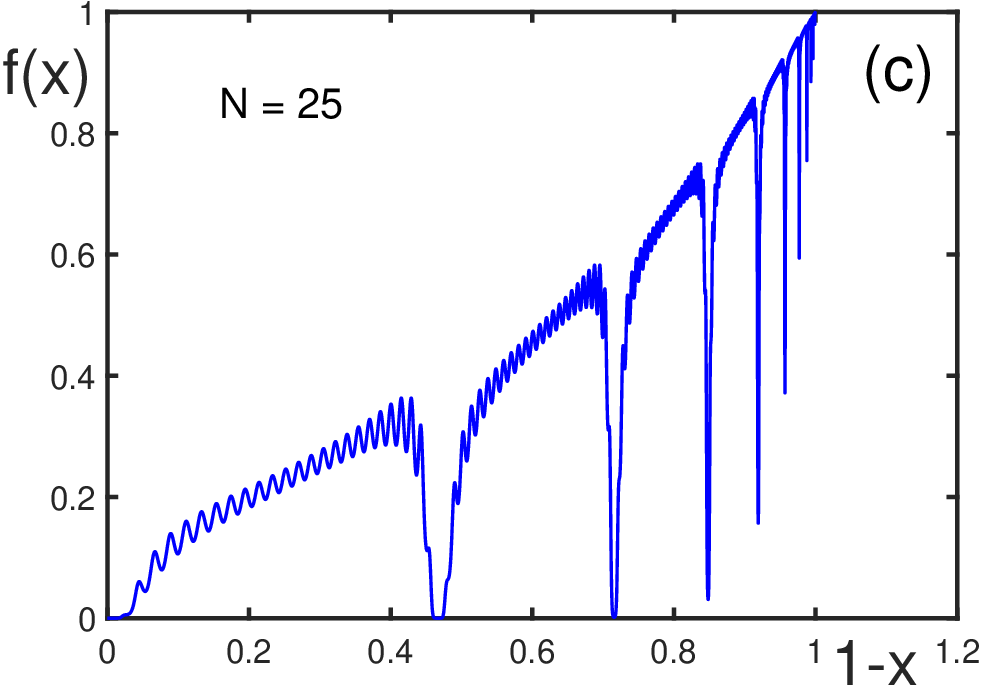} \hspace{1cm}
\includegraphics[width=7.5cm]{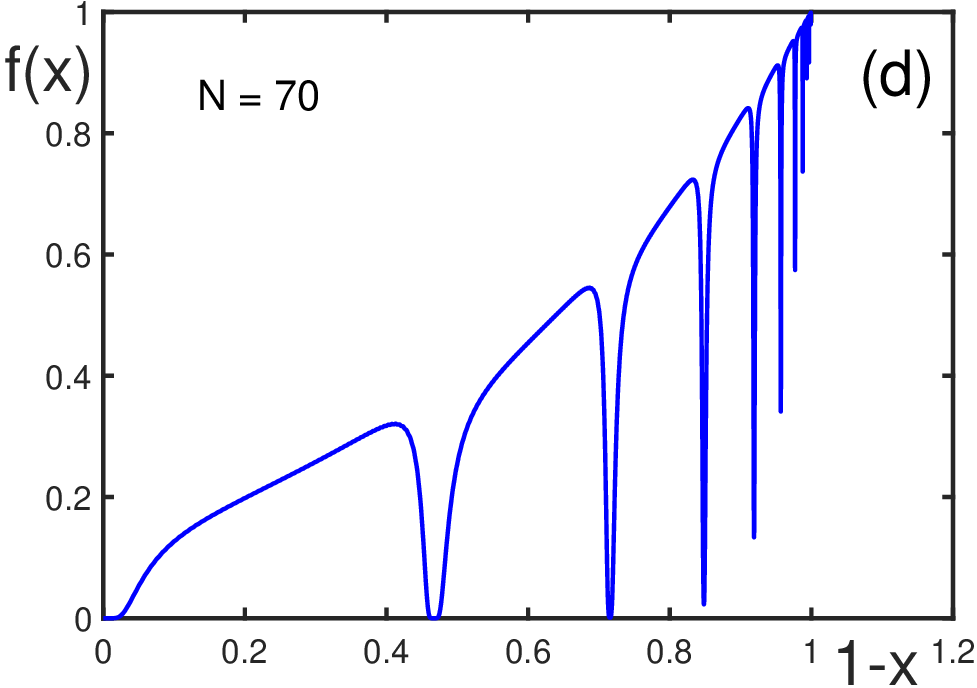} } }
\caption{\small
Function $f(x)$ for the fixed parameters $a=-3$, $b=-2$, $\al=1$, $\om=10$, $\mu=1.1$,
$\vp=0.1$, and $N$ varied:
(a) $N=1$;
(b) $N=5$;
(c) $N=25$;
(d) $N=70$.
Note that function $f(x)$ does not change its behaviour for $N > 70$.
}
\label{fig:Fig.6}
\end{figure}

\vskip 5cm

\begin{figure}[ht]
\centerline{
\hbox{
\includegraphics[width=7.5cm]{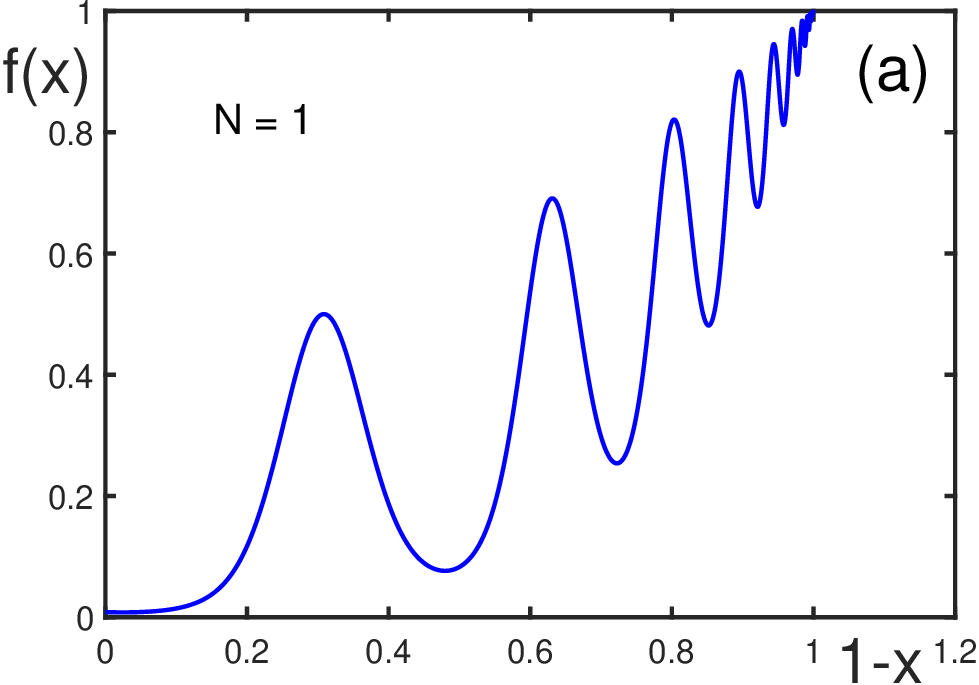} \hspace{1cm}
\includegraphics[width=7.5cm]{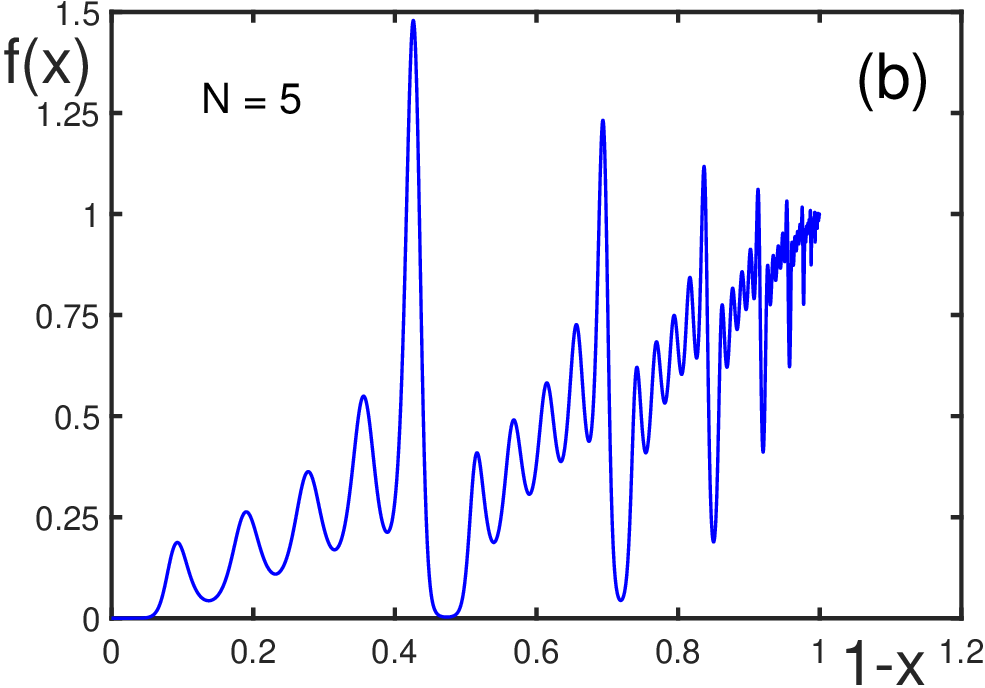} } }
\vskip 1cm
\centerline{
\hbox{
\includegraphics[width=7.5cm]{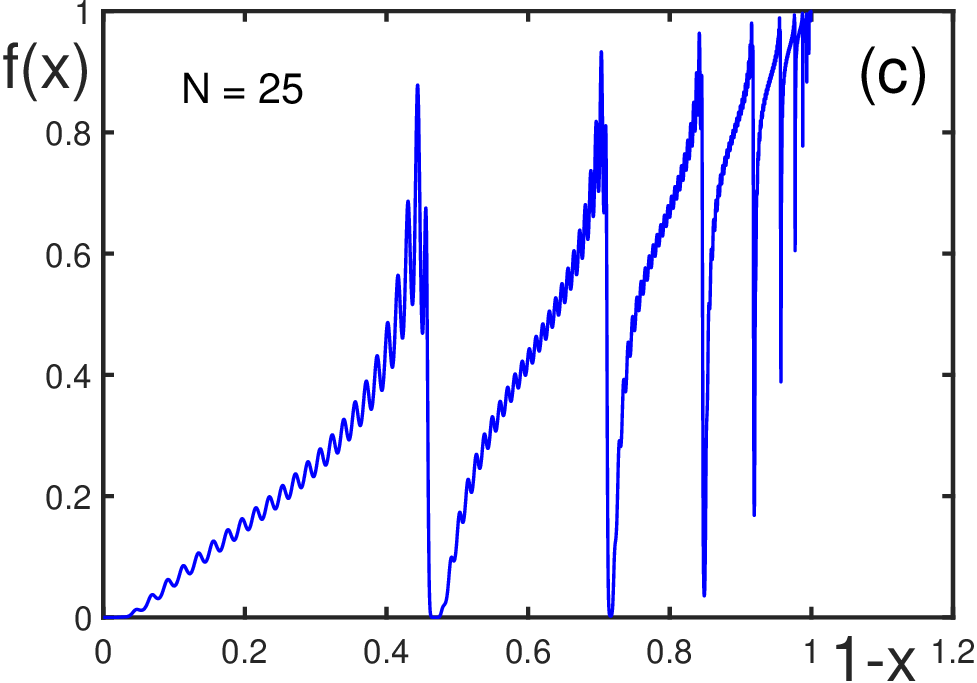} \hspace{1cm}
\includegraphics[width=7.5cm]{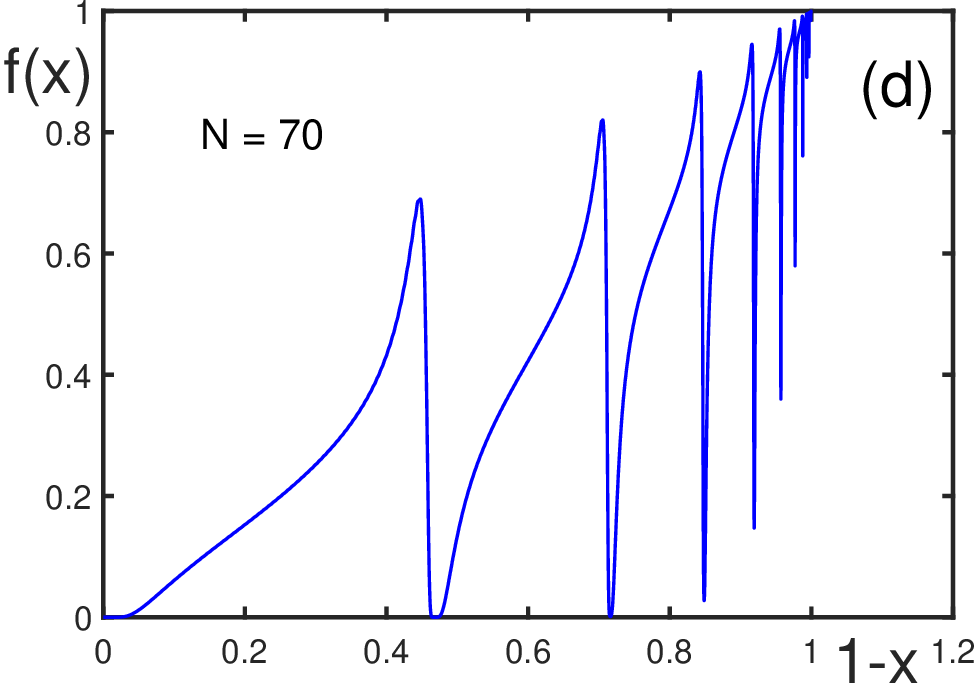} } }
\caption{\small
Function $f(x)$ for the fixed parameters $a=-3$, $b=-2$, $\al=1$, $\om=10$, $\mu=1.1$,
$\vp=0.5$, and $N$ varied:
(a) $N=1$;
(b) $N=5$;
(c) $N=25$;
(d) $N=70$.
Note that function $f(x)$ does not change its behaviour for $N > 70$.
}
\label{fig:Fig.7}
\end{figure}

\vskip 5cm

\begin{figure}[ht]
\centerline{
\hbox{
\includegraphics[width=7.5cm]{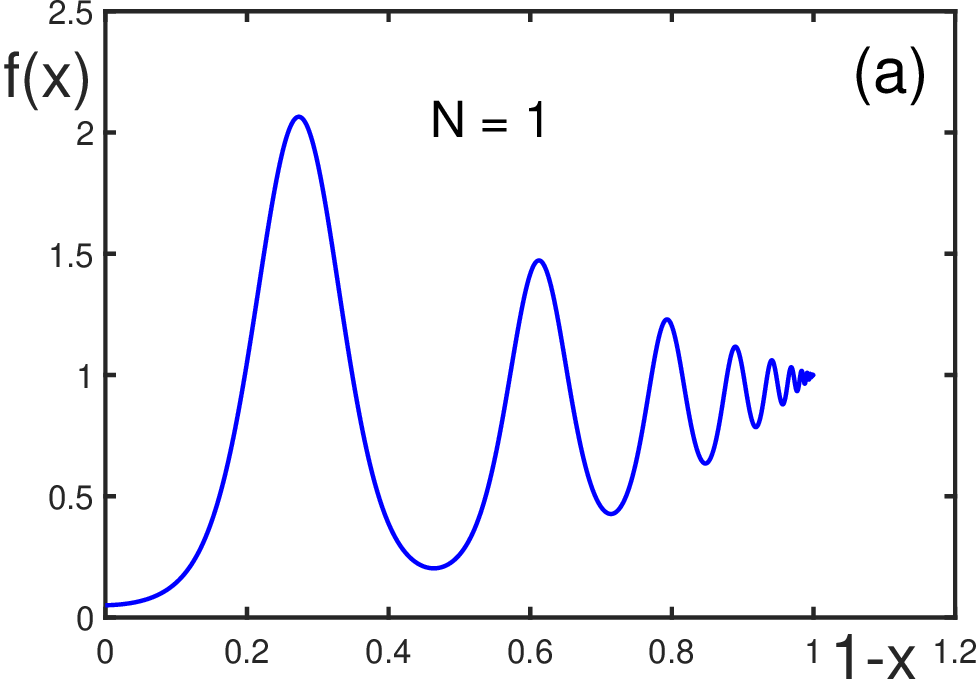} \hspace{1cm}
\includegraphics[width=7.5cm]{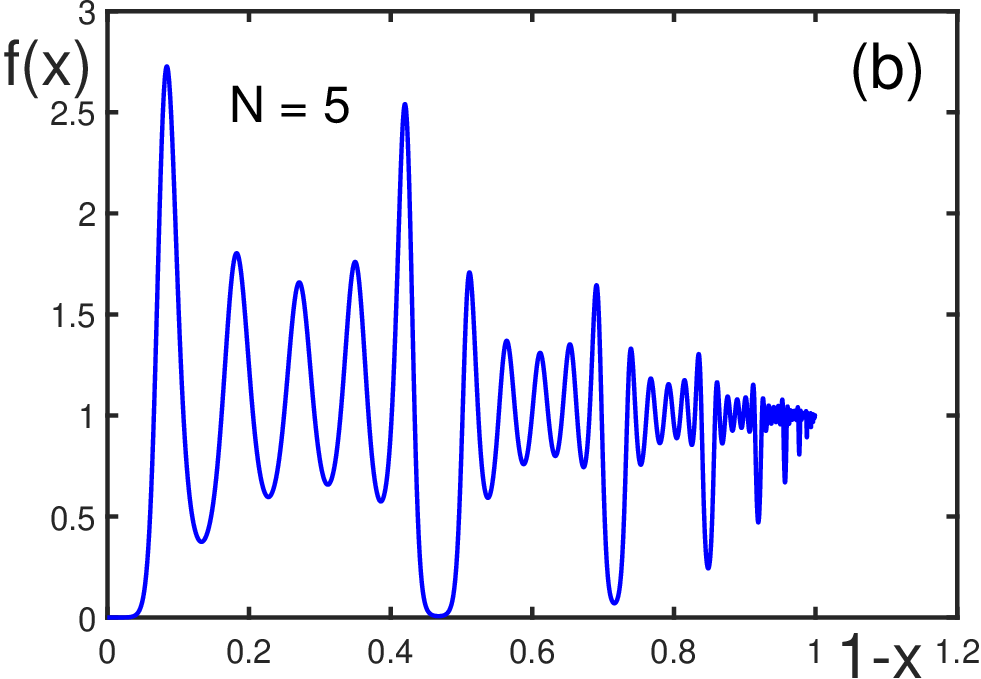} } }
\vskip 1cm
\centerline{
\hbox{
\includegraphics[width=7.5cm]{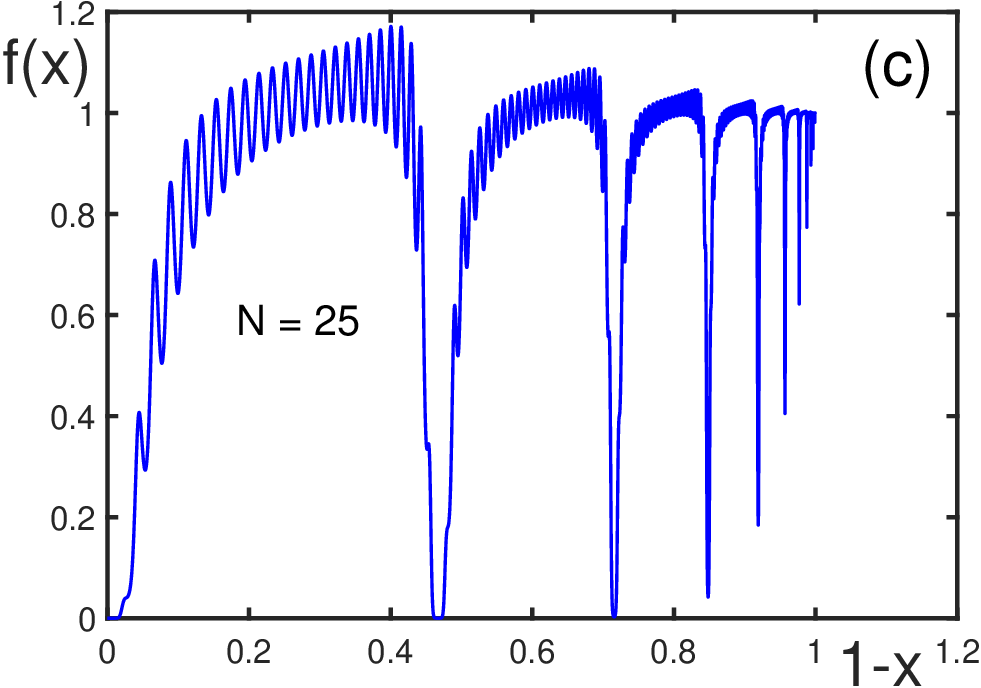} \hspace{1cm}
\includegraphics[width=7.5cm]{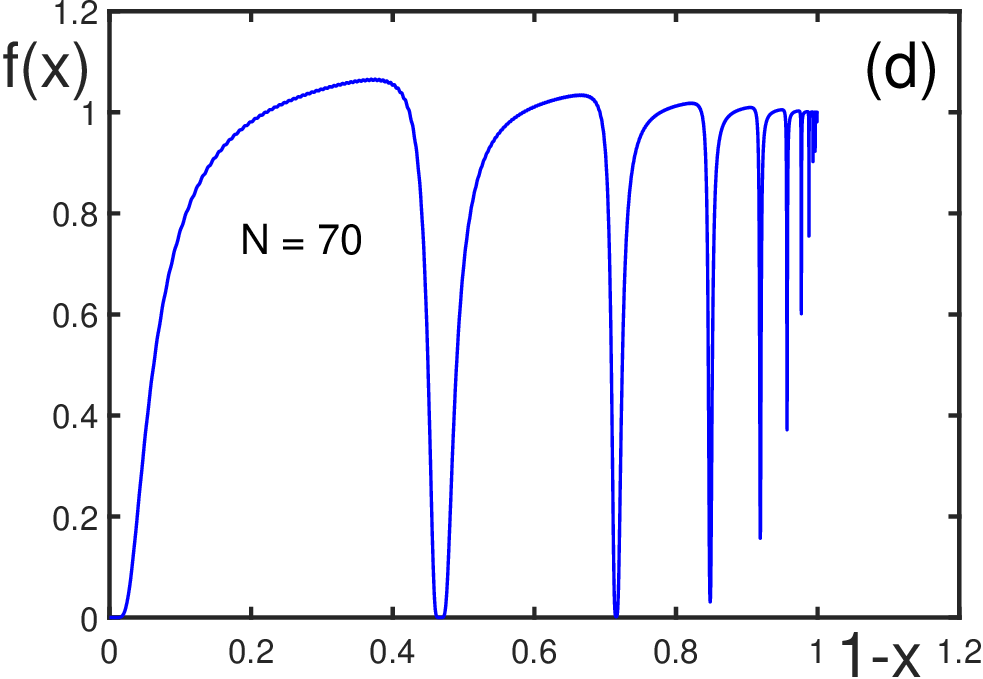} } }
\caption{\small
Function $f(x)$ for the fixed parameters $b=-2$, $\al=1$, $\om=10$, $\mu=1.1$, $\vp=0.1$,
$a=-1$, and $N$ varied:
(a) $N=1$;
(b) $N=5$;
(c) $N=25$;
(d) $N=70$.
Note that function $f(x)$ does not change its behaviour for $N > 70$.
}
\label{fig:Fig.8}
\end{figure}

\vskip 5cm

\begin{figure}[ht]
\centerline{
\hbox{
\includegraphics[width=7.5cm]{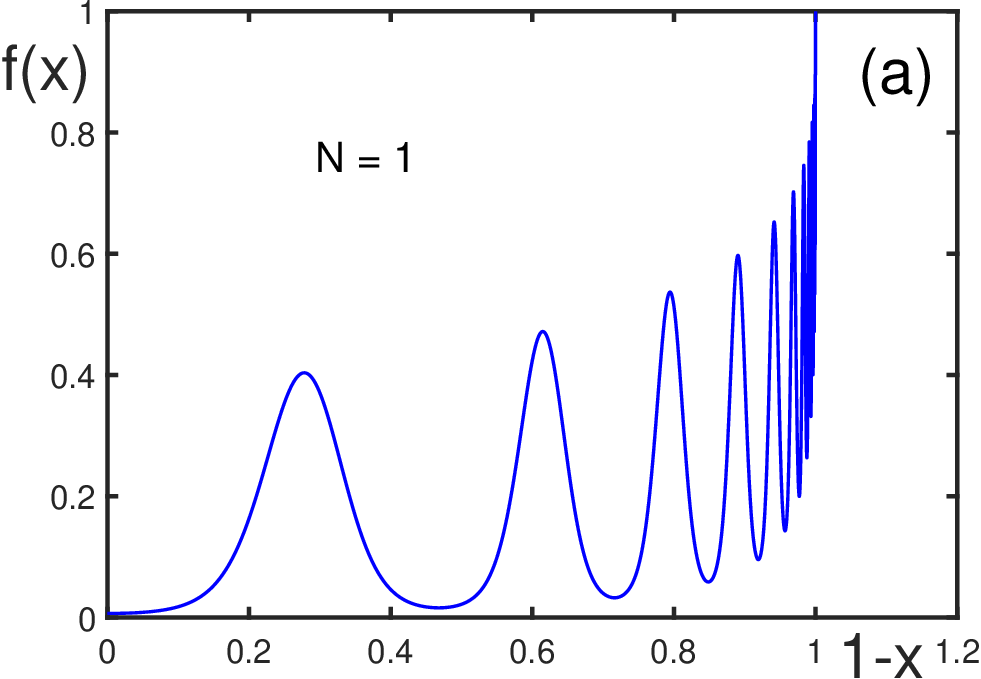} \hspace{1cm}
\includegraphics[width=7.5cm]{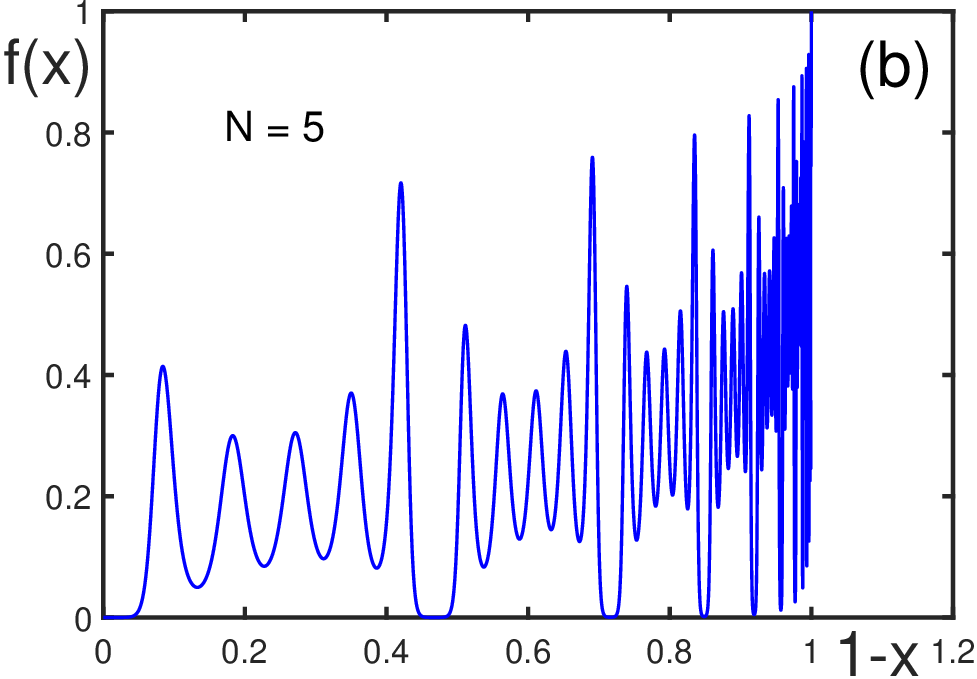} } }
\vskip 1cm
\centerline{
\hbox{
\includegraphics[width=7.5cm]{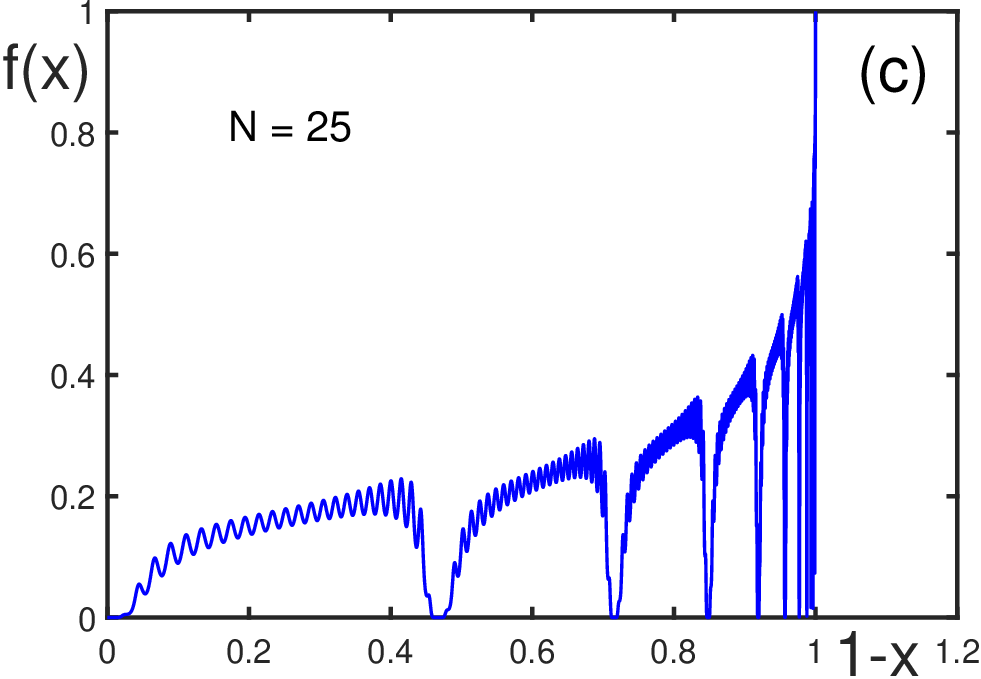} \hspace{1cm}
\includegraphics[width=7.5cm]{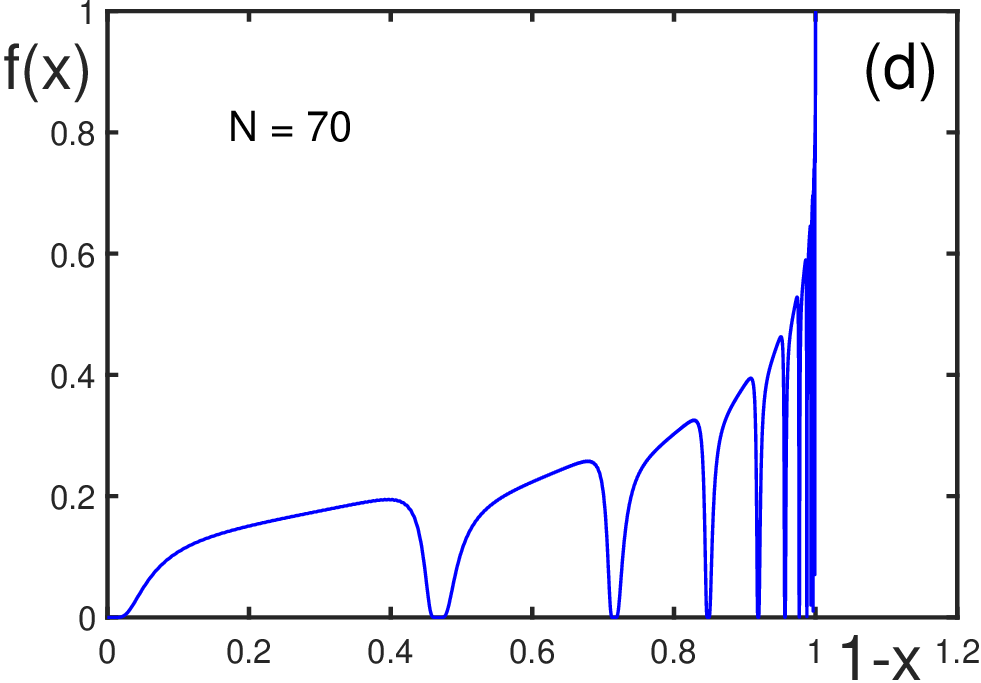} } }
\caption{\small
Function $f(x)$ for the fixed parameters $a=-3$, $b=-2$, $\om=10$, $\mu=1.1$,
$\vp=0.1$, $\al=0.3$, and $N$ varied:
(a) $N=1$;
(b) $N=5$;
(c) $N=25$;
(d) $N=70$.
Note that function $f(x)$ does not change its behaviour for $N > 70$.
}
\label{fig:Fig.9}
\end{figure}

\vskip 5cm

\begin{figure}[ht]
\centerline{
\hbox{
\includegraphics[width=7.5cm]{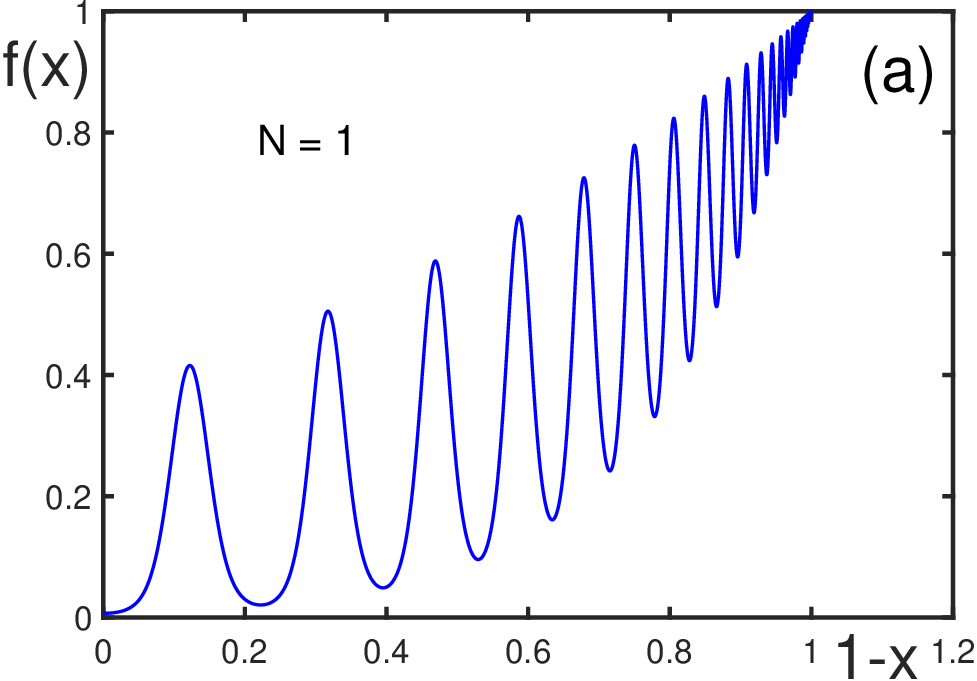} \hspace{1cm}
\includegraphics[width=7.5cm]{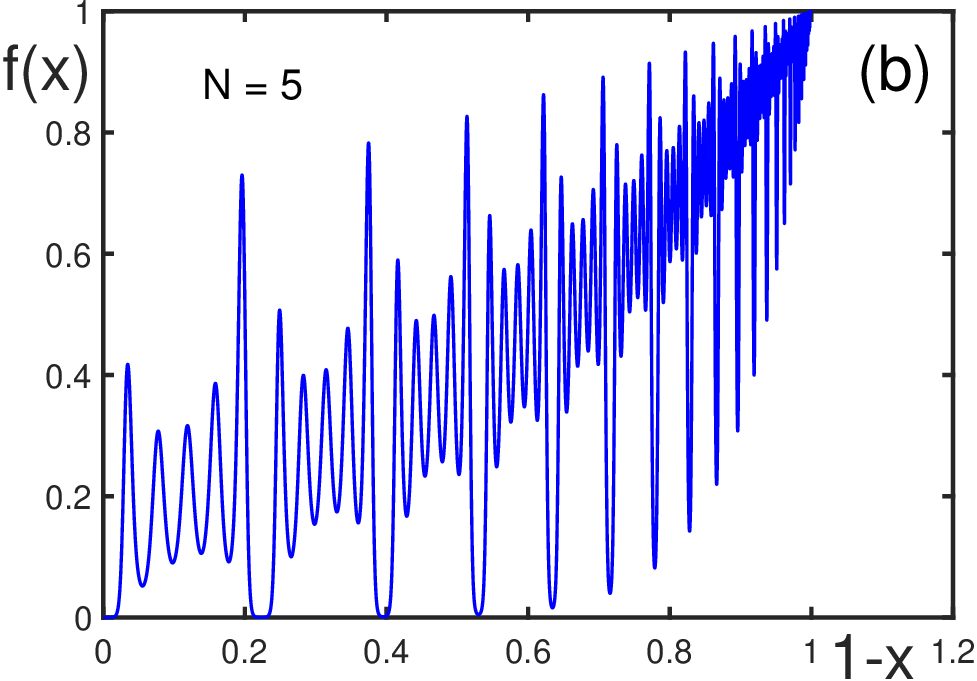} } }
\vskip 1cm
\centerline{
\hbox{
\includegraphics[width=7.5cm]{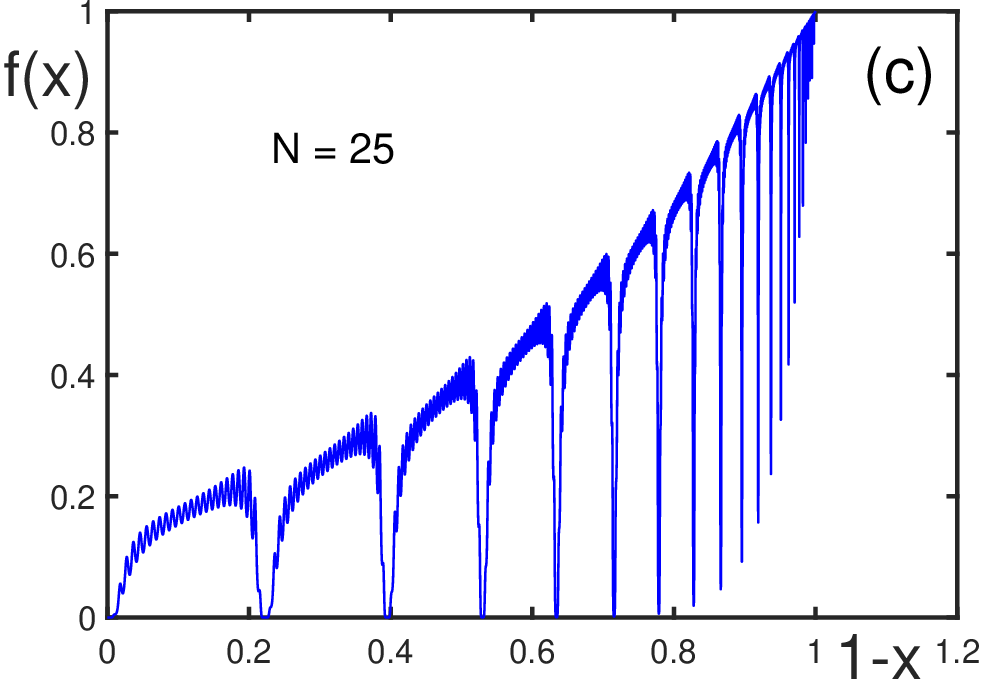} \hspace{1cm}
\includegraphics[width=7.5cm]{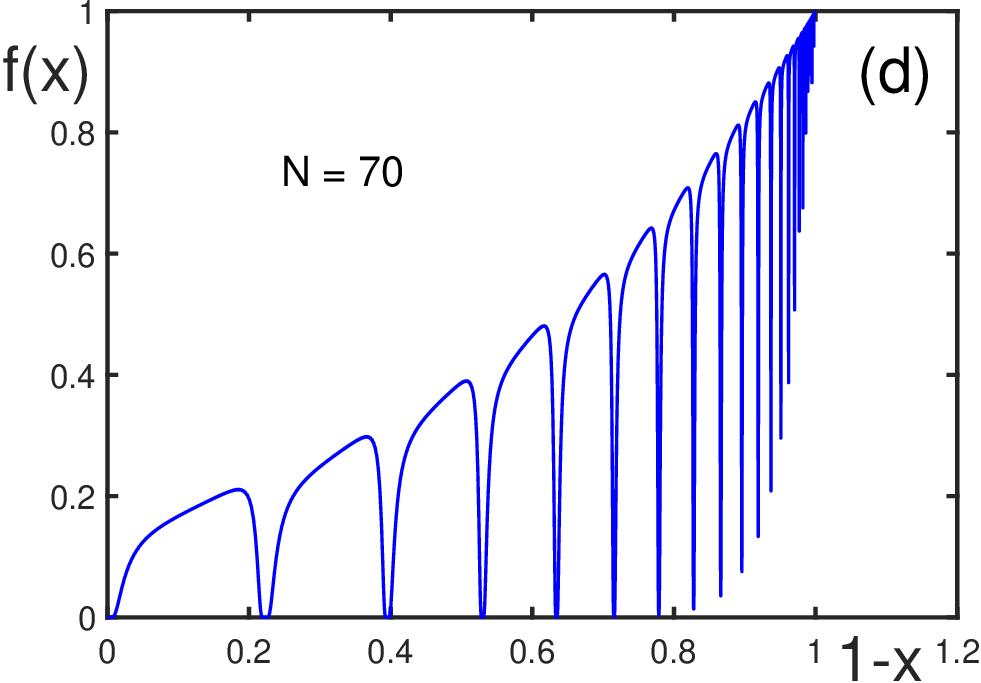} } }
\caption{\small
Function $f(x)$ for the fixed parameters $a=-3$, $b=-2$, $\mu=1.1$, $\vp=0.1$, $\al=1$,
$\om=25$, and $N$ varied:
(a) $N=1$;
(b) $N=5$;
(c) $N=25$;
(d) $N=70$.
Note that function $f(x)$ does not change its behaviour for $N > 70$.
}
\label{fig:Fig.10}
\end{figure}

\vskip 5cm

\begin{figure}[ht]
\centerline{
\hbox{
\includegraphics[width=7.5cm]{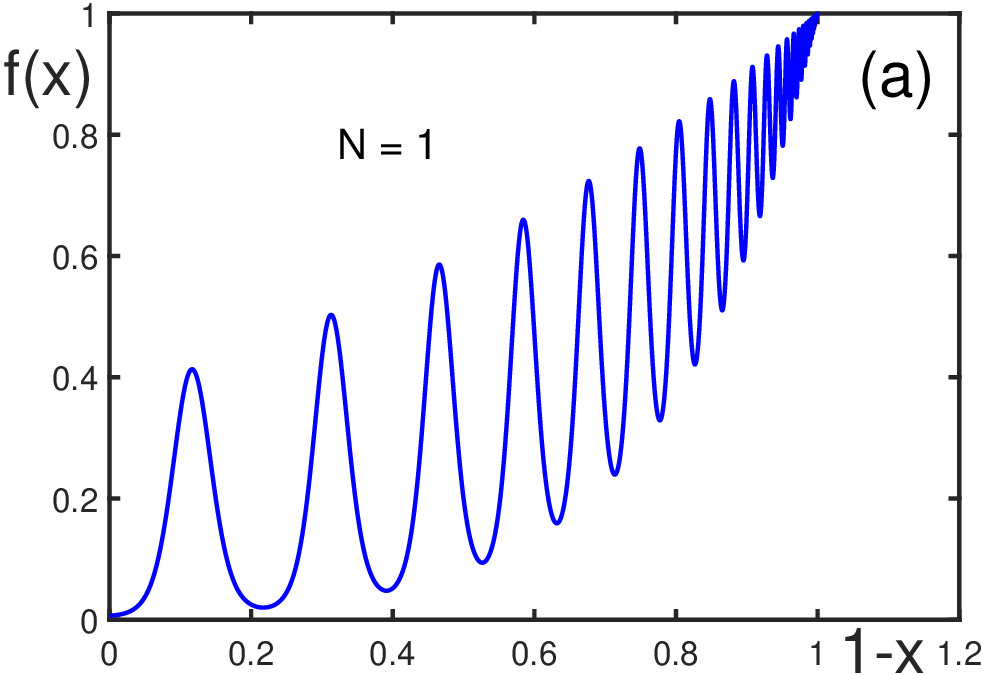} \hspace{1cm}
\includegraphics[width=7.5cm]{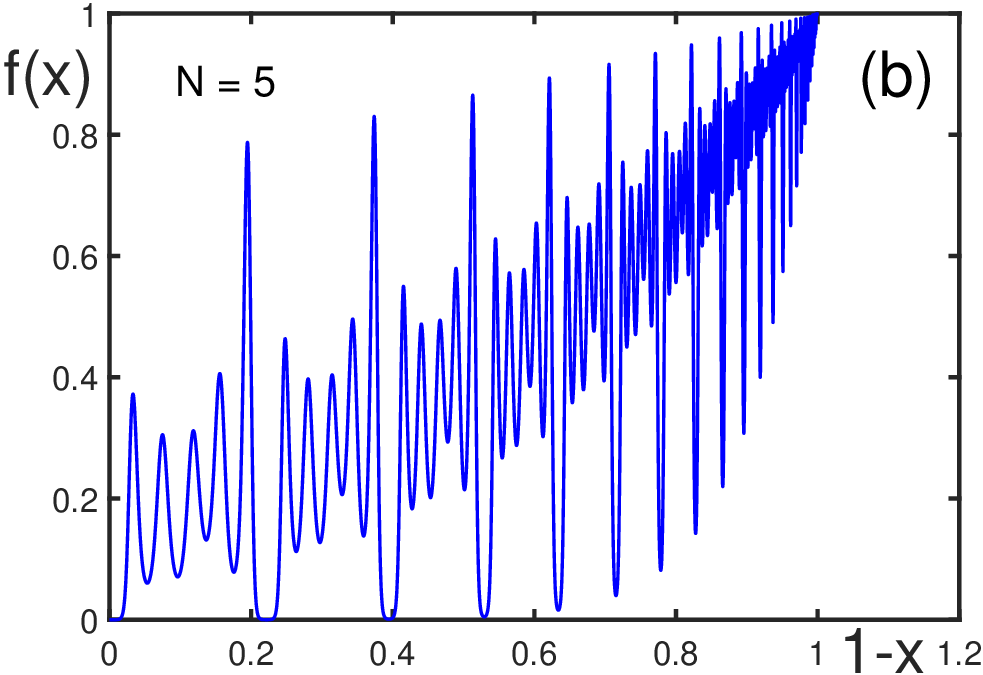} } }
\vskip 1cm
\centerline{
\hbox{
\includegraphics[width=7.5cm]{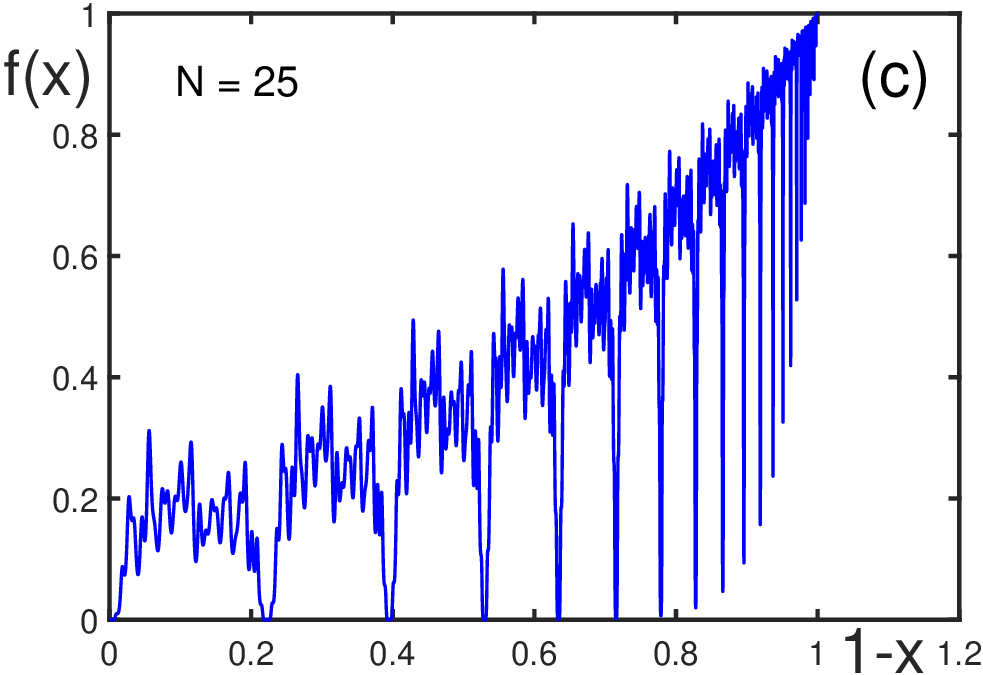} \hspace{1cm}
\includegraphics[width=7.5cm]{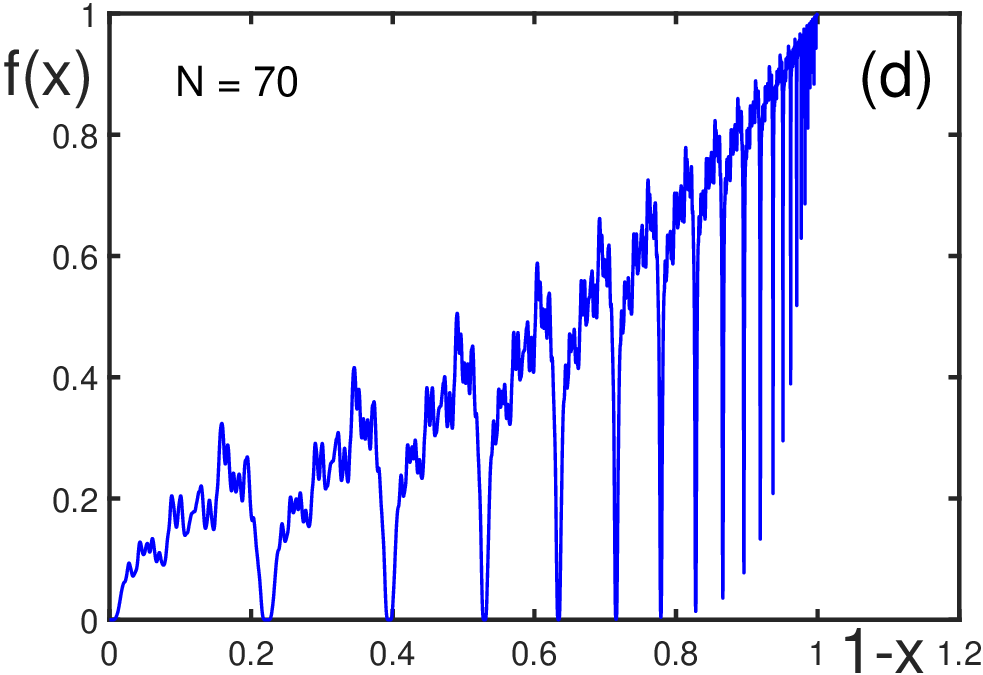} } }
\vskip 1cm
\centerline{
\hbox{
\includegraphics[width=7.5cm]{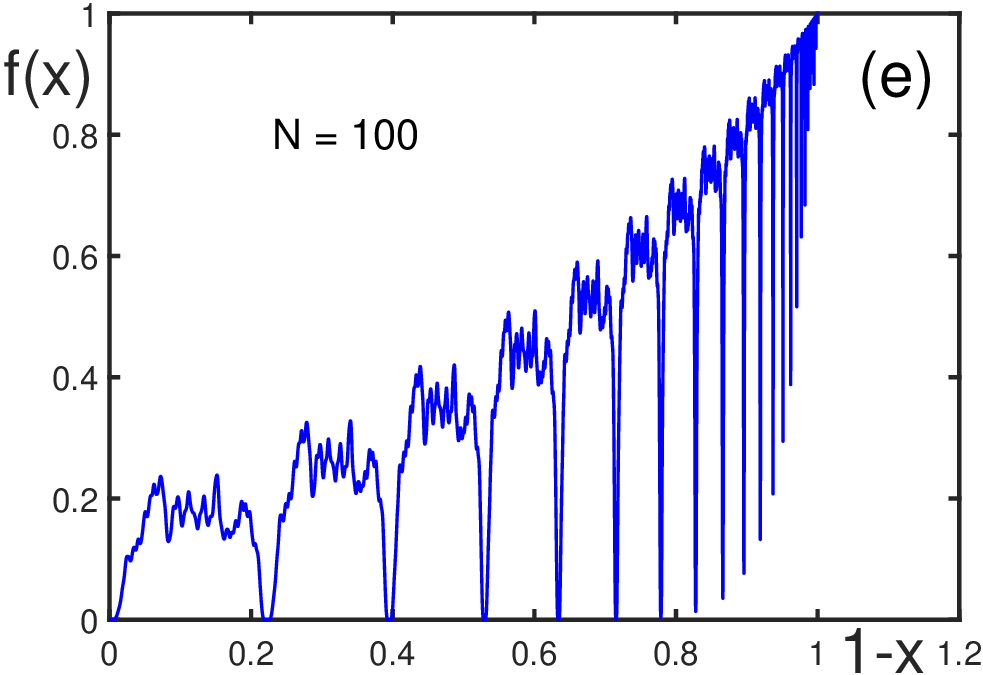} \hspace{1cm}
\includegraphics[width=7.5cm]{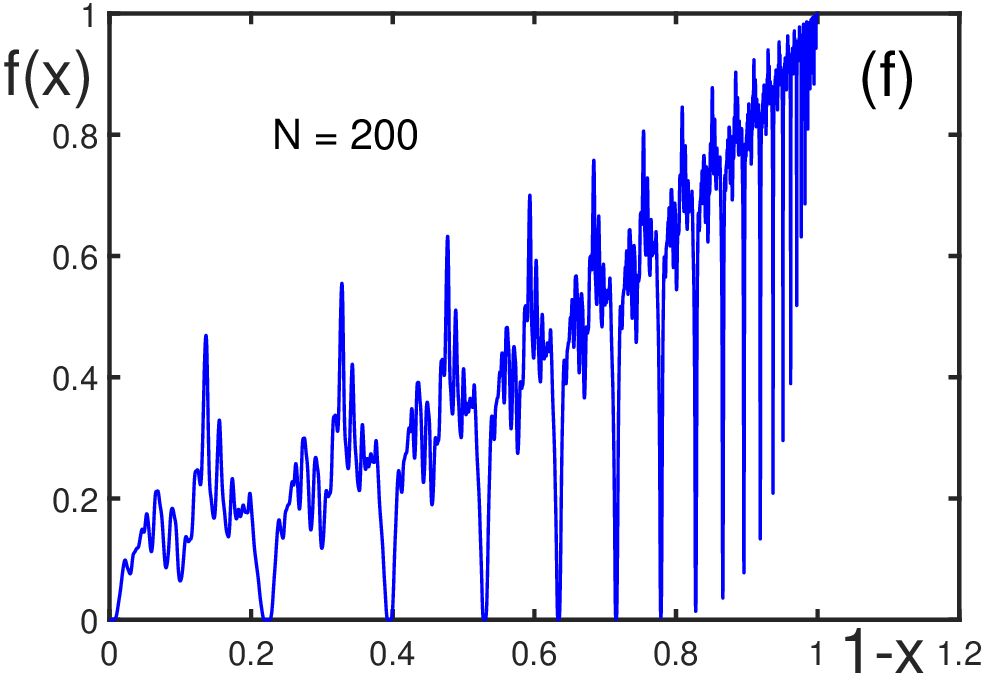} } }
\caption{\small
Function $f(x)$ for the fixed parameters $a=-3$, $b=-2$, $\om=25$, $\al=1$,
$\mu=1.1$, and $N$ varied: 
(a) $N=1$; 
(b) $N=5$; 
(c) $N=25$; 
(d) $N=70$;
(e) $N=100$; 
(f) $N=200$. 
The parameters $\vp_n$ are generated as random numbers from the normal distribution with 
standard deviation $\sgm=0.1$ and zero mean.
}
\label{fig:Fig.11}
\end{figure}

\vskip 5cm

\begin{figure}[ht]
\centerline{
\hbox{
\includegraphics[width=7.5cm]{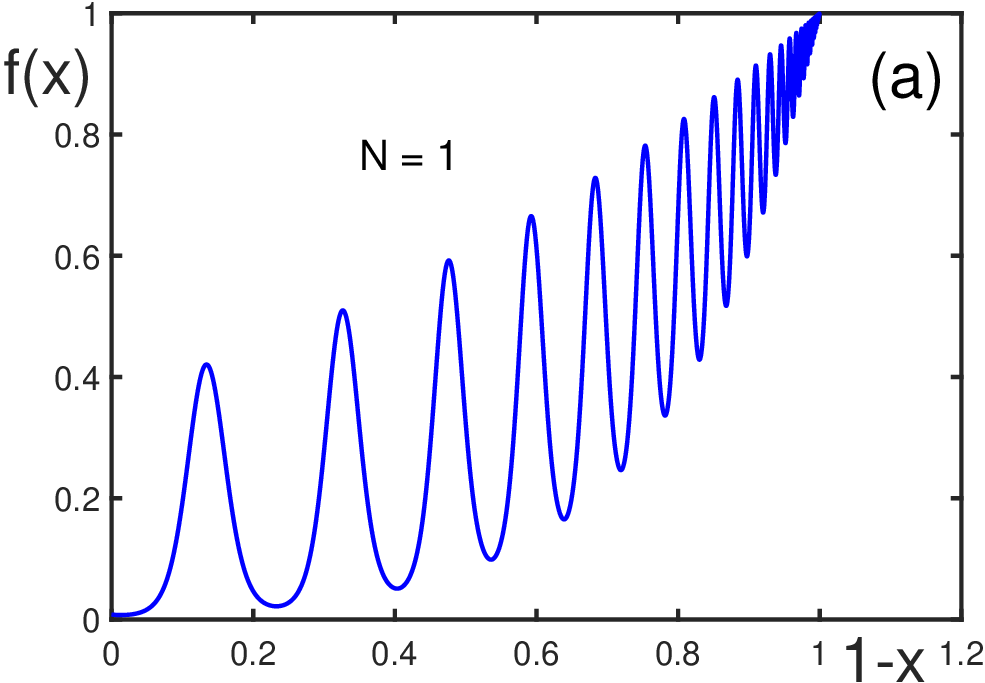} \hspace{1cm}
\includegraphics[width=7.5cm]{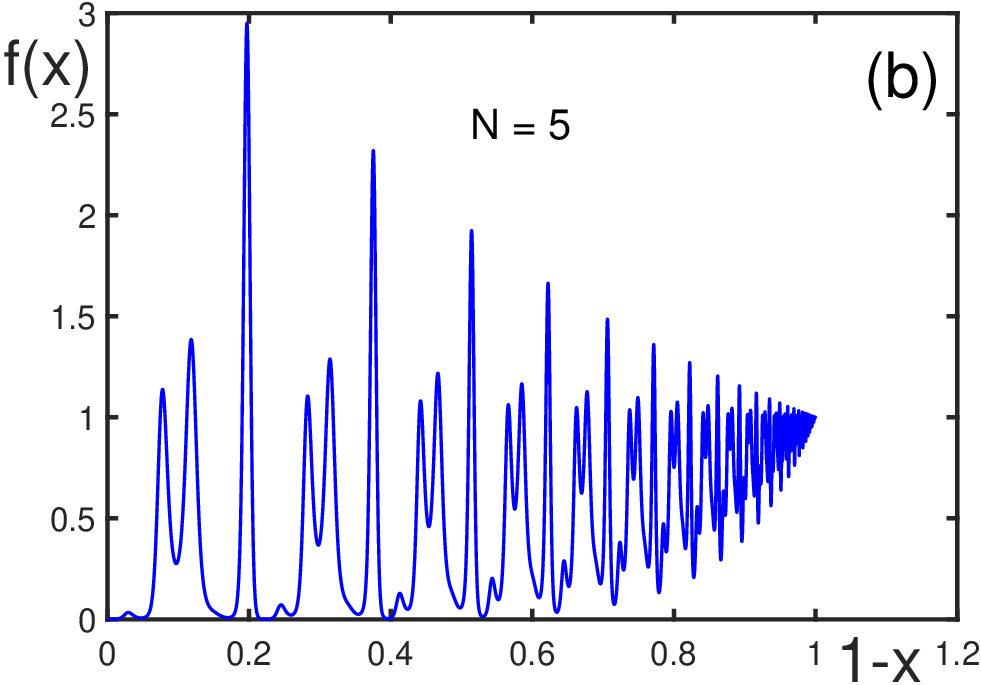} } }
\vskip 1cm
\centerline{
\hbox{
\includegraphics[width=7.5cm]{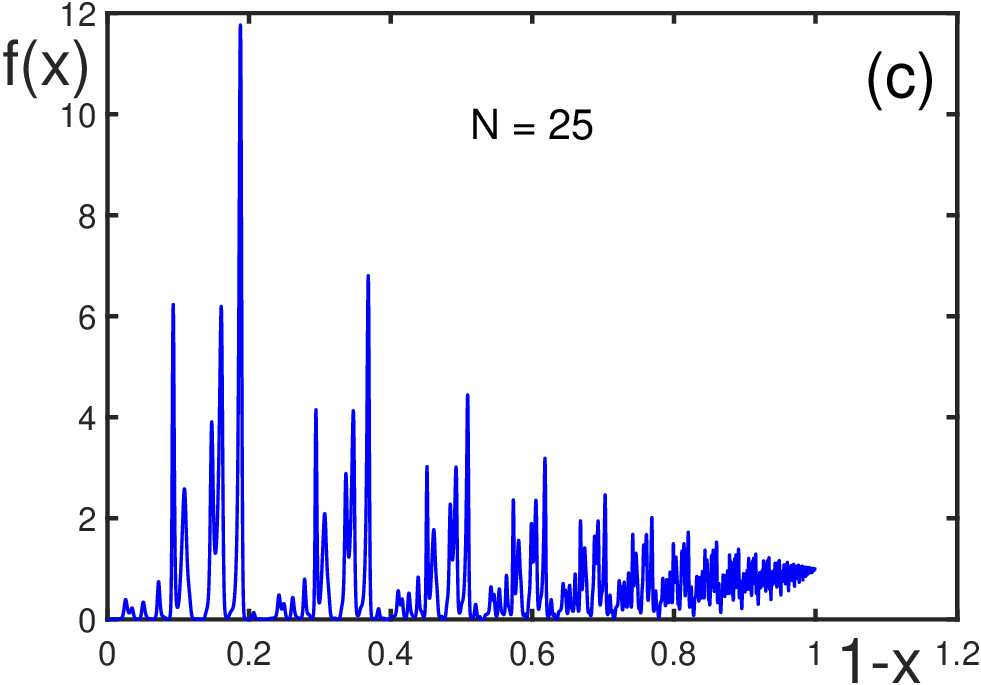} \hspace{1cm}
\includegraphics[width=7.5cm]{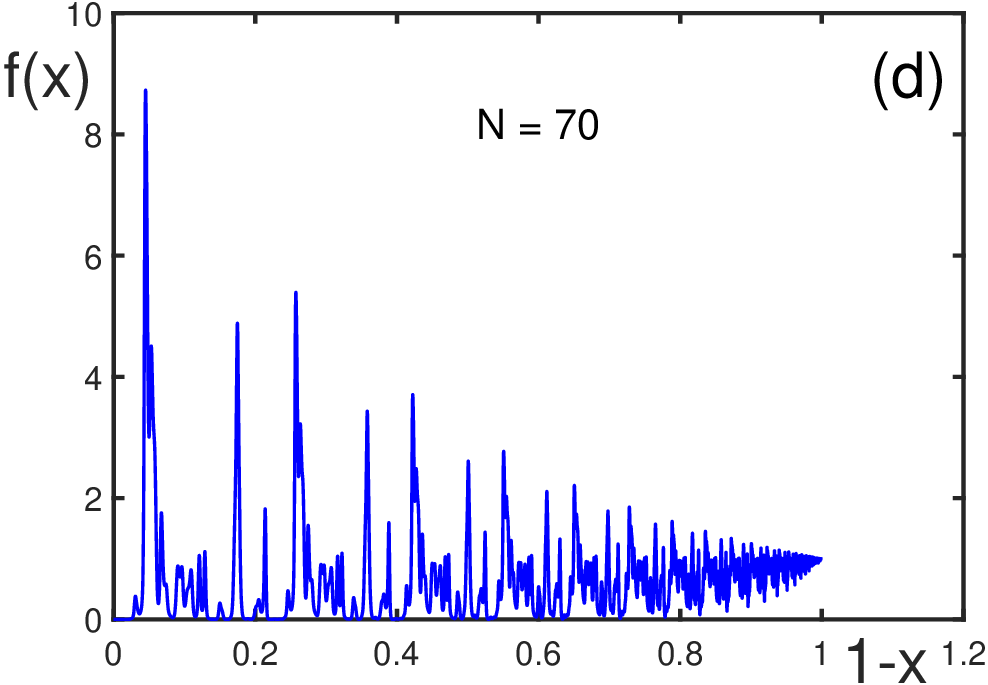} } }
\vskip 1cm
\centerline{
\hbox{
\includegraphics[width=7.5cm]{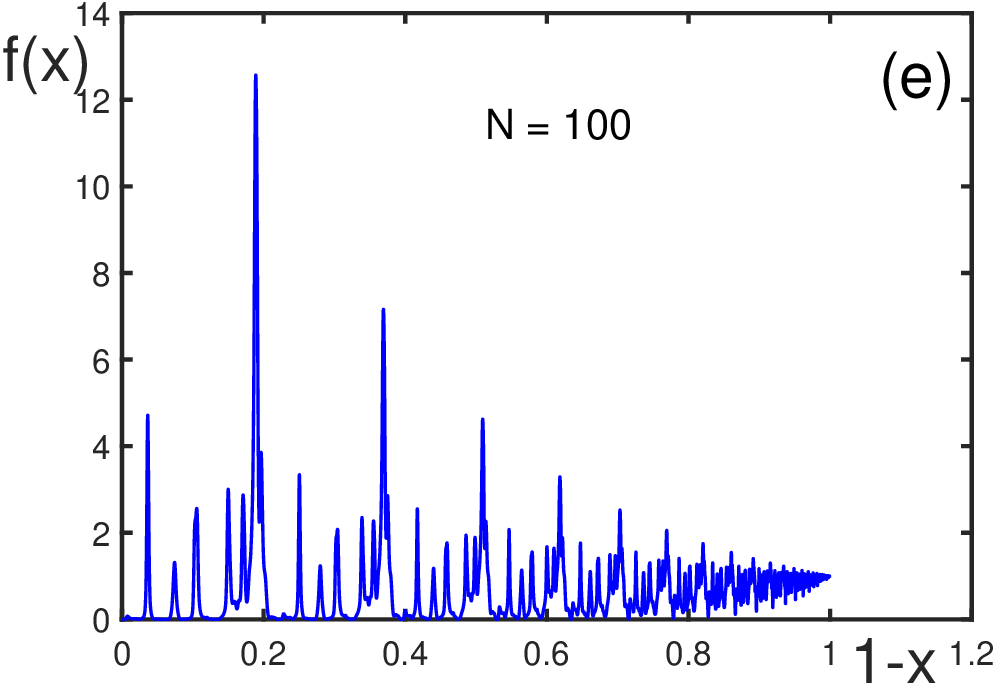} \hspace{1cm}
\includegraphics[width=7.5cm]{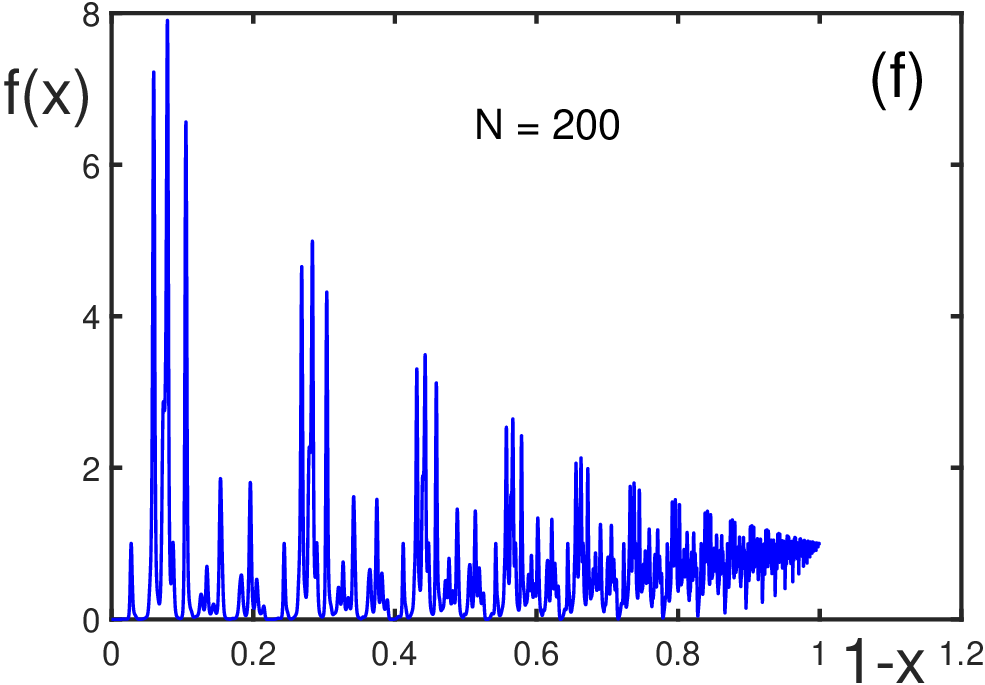} } }
\caption{\small
Function $f(x)$ for the fixed parameters $a=-3$, $b=-2$, $\om=25$, $\al=1$,
$\mu=1.1$, and $N$ varied: 
(a) $N=1$; 
(b) $N=5$; 
(c) $N=25$; 
(d) $N=70$;
(e) $N=100$; 
(f) $N=200$. 
The parameters $\vp_n$ are generated as random numbers
from the normal distribution with standard deviation $\sgm=1$ and zero mean.
}
\label{fig:Fig.12}
\end{figure}


\newpage

\begin{thebibliography}{99}

\bibitem{Fisher_1}
Fisher, M.E.
{\it The Nature of Critical Points};
University of Colorado: Boulder, 1965.

\bibitem{Brout_2}
Brout, R.
{\it Phase Transitions};
Benjamin: New York, 1965.

\bibitem{Stanley_3}
Stanley, H.E.
{\it Introduction to Phase Transitions and Critical Phenomena};
Oxford University: Oxford, 1987.

\bibitem{Yukalov_4}
Yukalov, V.I.; Shumovsky, A.S.
{\it Lectures on Phase Transitions};
World Scientific: Singapore, 1990.

\bibitem{Sornette_5}
Sornette, D.
{\it Critical Phenomena in Natural Sciences};
Springer: Berlin, 2006.

\bibitem{Bogolubov_6}
Bogolubov, N.N.
{\it Quantum Statistical Mechanics};
World Scientific, Singapore, 2015.

\bibitem{Wilson_7}
Wilson, K.G.; Kogut, J.
The renormalization group and the $\varepsilon$ expansion.
{\it Phys. Rep.} {\bf 1974}, {\it 12}, 75--199. 

\bibitem{Hu_8}
Hu, B.
Introduction to real-space renormalization group methods in critical and chaotic 
phenomena. 
{\it Phys. Rep.} {\bf 1982}, {\it 91}, 233--295. 

\bibitem{Bogolubov_9}
Bogolubov, N.N.; Shirkov, D.V.
{\it Quantum Fields};
Benjamin: London, 1983. 

\bibitem{Kadanoff_10}
Kadanoff, L.P.
{\it Statistical Physics: Statics, Dynamics and Renormalization};
World Scientific: Singapore, 2000. 

\bibitem{Ma_11}
Ma, S.K.
{\it Modern Theory of Critical Phenomena};
Routledge: New York, 2009.

\bibitem{Efrati_12}
Efrati, E.; Wang, Z.; Kolan, A.; Kadanoff, L.P.
Real-space renormalization in statistical mechanics.
{\it Rev. Mod. Phys.} {\bf 2014}, {\it 86}, 647--669. 

\bibitem{Yukalov_13}
Yukalov, V.I.
Interplay between approximation theory and renormalization group.
{\it Phys. Part. Nucl.} {\bf 2019}, {\it 50}, 141--209. 

\bibitem{Dupuis_27}
Dupuis, N.; Canet, L.; Eichhorn, A.; Metzner, W.; Pawlowski, J.M. Tissier, M.; Wschebor, N.
The nonperturbative functional renormalization group and its applications.
{\it Phys. Rep.} {\bf 2021}, {\it 910}, 1--114. 

\bibitem{Sornette_14}
Sornette, D.
Discrete scale invariance and complex dimensions.
{\it Phys. Rep.} {\bf 1998}, {\it 297}, 239--270.
(extended version at http://xxx.lanl.gov/abs/cond-mat/9707012)


\bibitem{Harris_1948}
Harris, T.E.
Branching processes.
{\it Ann. Math. Stat.}, {\bf 1948}, {\it 19}, 474--494.

\bibitem{Derrida}
Derrida, B.; Eckmann, J.P.; Erzan, A
Renormalisation groups with periodic and aperiodic orbits.
{\it J. Phys. A} {\bf 1983}, {\it 16}, 893--906.

\bibitem{Derrida_2}
Derrida, B.; De Seze, L.; Itzykson C.
Fractal structure of zeros in hierarchical models.
{\it J. Stat. Phys.} {\bf 1983}, {\it 33}, 559--569.

\bibitem{Derrida_3}
Derrida, B.; Itzykson, C.; Luck, J.M.
Oscillatory critical amplitudes in hierarchical models.
{\it Comm. Math. Phys.} {\bf 1984}, {\it 94}, 115--132.

\bibitem{Bessis}
Bessis, D.; Geronimo, J.S.; Moussa, P.
Mellin transforms associated with Julia sets and physical applications.
{\it J. Stat. Phys.} {\bf 1984}, {\it 34}, 75--110.

\bibitem{Itzykson}
Itzykson, C.; Luck, J.M.
Zeroes of the partition function for statistical models on regular and hierarchical lattices.
{\it Prog. Phys.} {\bf 1985}, {\it 11}, 45--82.

\bibitem{Costin}
Costin, O.; Giacomin, G.
Oscillatory critical amplitudes in hierarchical models and the Harris function of branching 
processes.
{\it J. Stat. Phys.} {\bf 2013}, {\it 150}, 471--486.

\bibitem{Derrida_4}
Derrida, B.; Giacomin, G.
Log-periodic critical amplitudes: A perturbative approach.
{\it J. Stat. Phys.} {\bf 2014}, {\it 154}, 286--304.


\bibitem{Vallejos}
Vallejos, R.O.; Anteneodo, C.
Thermodynamical fingerprints of fractal spectra.
{\it Phys. Rev. E} {\bf 1998} {\it 58}, 4134--4140.

\bibitem{Knezevic}
Knezevic, M.; Knezevic, D.
Oscillatory behavior of critical amplitudes of the Gaussian model on a hierarchical structure
{\it Phys. Rev. E} {\bf 1999}, {\it 60}, 3396--3398.

\bibitem{Lessa}
Lessa, J.C.; Andrade, R.F.S.
Log-periodic oscillations for a uniform spin model on a fractal.
{\it Phys. Rev. E} {\bf 2000}, {\it 62}, 3083--3089.

\bibitem{Bab}
Bab, M.A.; Fabricius, G.; Albano, E.V.
On the occurrence of oscillatory modulations in the power law behavior of dynamic and 
kinetic processes in fractals.
{\it Europhys. Lett.} {\bf 2008}, {\it 81}, 10003.

\bibitem{Padilla}
Padilla, L.; Martin, H.O.; Iguain, J.L.
Log-periodic modulation in one-dimensional random walks.
{\it Europhys. Lett.} {\bf 2009}, {\it 85}, 20008.

\bibitem{Akkermans}
Akkermans, E.; Benichou, O.; Dunne, G.V.; Teplyaev, A.; Voituriez R.
Spatial log-periodic oscillations of first-passage observables in fractals.
{\it Phys. Rev. E} {\bf 2012}, {\it 86}, 061125.

\bibitem{Dunne}
Dunne, G.V.
Heat kernels and zeta functions on fractals.
{\it J. Phys. A} {\bf 2012}, {\it 45}, 374016.


\bibitem{Luck}
Luck, J.M.; Nieuwenhuizen, T.M.
A soluble quasi-crystalline magnetic model: The XY quantum spin chain.
{\it Europhys. Lett.} {\bf 1986}, {\it 2}, 257--266.

\bibitem{Karevski}
Karevski, D.; Turban, L.
Log-periodic corrections to scaling: Exact results for aperiodic Ising quantum chains.
{\it J. Phys. A} {\bf 1996}, {\it 29}, 3461--3470.

\bibitem{Andrade}
Andrade, R.F.S.
Detailed characterization of log-periodic oscillations for an aperiodic Ising model.
{\it Phys. Rev. E} {\bf 2000}, {\it 61}, 7196--7199.

\bibitem{Carpena}
Carpena, P.; Coronado, A.V.; Bernaola-Galvan, P.
Thermodynamics of fractal spectra: Cantor sets and quasiperiodic sequences.
{\it Phys. Rev. E} {\bf 2000}, {\it 61}, 2281--2289.

\bibitem{Johansen-Sammis-95}
Sornette, D.; Sammis, C.G.
Complex critical exponents from renormalization group theory of earthquakes:
Implications for earthquake predictions.
{\it J. Phys. I France} {\bf 1995} {\it 5}, 607--619. 

\bibitem{Johansen-wakita_96}
Johansen, A.; Sornette, D.; Wakita, H.; Tsunogai, U.; Newman, W.I.; Saleur, H.
Discrete scaling in earthquake precursory phenomena: Evidence in the Kobe earthquake, Japan.
{\it J. Phys. I France} {\bf 1996}, {\it 6}, 1391--1402 .

\bibitem{Anifrani-sor-95}
Anifrani, J.C.; Le Floc'H, C.; Sornette, D.; Souillard, B.
Universal Log-periodic correction to renormalization group scaling for rupture stress
prediction from acoustic emissions,
{\it J. Phys. I France} {\bf 1995}, {\it 5}, 631--638. 

\bibitem{Johansen_sor-1998}
Johansen, A.; Sornette, D.
Evidence of discrete scale invariance by canonical averaging.
{\it Int. J. Mod. Phys. C} {\bf 1998}, {\it 9}, 433--447. 

\bibitem{Johansen_15}
Johansen, A.; Sornette, D.
Critical ruptures,
{\it Eur. Phys. J. B} {\bf 2000}, {\it 18}, 163--181. 

\bibitem{Moura_16}
Moura, A.; Yukalov, V.I.
Self-similar extrapolation for the law of acoustic emission before failure of
heterogeneous materials.
{\it Int. J. Fract.} {\bf 2002}, {\it 118}, 63--68. 

\bibitem{Yukalov_17}
Yukalov, V.I.; Moura, A.; Nechad, H.
Self-similar law of energy release before materials fracture.
{\it J. Mech. Phys. Solids} {\bf 2004}, {\it 52}, 453--465. 


\bibitem{Sornette_bou96}
Sornette, D.; Johansen, A.; Bouchaud, J.-P.
Stock market crashes, precursors and replicas.
{\it J. Phys. I France} {\bf 1996}, {\it 6}, 167--175. 

\bibitem{Sornette_18}
Sornette, D.; Johansen, A.
Large financial crashes.
{\it Physica A} {\bf 1997}, {\it 245}, 411--422. 

\bibitem{Johansen_19}
Johansen, A.; Sornette, D.
Modeling the stock market prior to large crashes.
{\it Eur. Phys. J. B} {\bf 1999}, {\it 9}, 167--174. 

\bibitem{Johansen_20}
Johansen, A.; Sornette, D.
Bubbles and anti-bubbles in Latin-American, Asian and western markets: An empirical study.
{\it Int. J. Theor. Appl. Finance} {\bf 2001}, {\it 4}, 853--920. 

\bibitem{Sornette_21}
Sornette, D.; Johansen, A.
Significance of log-periodic precursors to financial crashes.
{\it Quantit. Finance} {\bf 2001}, {\it 1}, 452--471. 

\bibitem{Feigenbaum_2001}
Feigenbaum, J.A. 
A statistical analysis of log-periodic precursors to financial crashes.
{\it Quantit. Finance} {\bf 2001}, {\it 1}, 346--360.

\bibitem{sorn-book-crash-02}
Sornette, D.
{\it Why Stock Markets Crash (Critical Events in Complex Financial Systems)};
Princeton University Press: Princeton, 2002; second print with extended preface, 2017.

\bibitem{Feigenbaum_22}
Feigenbaum, J.
Financial physics.
{\it Rep. Prog. Phys.} {\bf 2003}, {\it 66}, 1611--1649. 

\bibitem{Clark_2004}
Clark, A.  
Evidence of log-periodicity in corporate bond spreads. 
{\it Physica A} {\bf 2004}, {\it 338}, 585--595.


\bibitem{Bree_2013}
Bree, D.S.; Joseph, N.L.  
Testing for financial crashes using the log periodic power law model. 
{\it Int. Rev. Financial Anal.} {\bf 2013}, {\it 30}, 287--297.

\bibitem{Fantazzini_2013}
Fantazzini, D.; Geraskin, P.  
Everything you always wanted to know about log periodic power laws for bubble modelling 
but were afraid to ask. 
{\it Eur. J. Finance} {\bf 2013}, {\it 19}, 366--392.

\bibitem{Gustavsson_2016}
Gustavsson, M.; Leven, D.; Sjogren, H.
The timing of the popping: Using the log-periodic power law model to predict the bursting 
of bubbles on financial markets.
{\it Financ. Hist. Rev.} {\bf 2016}, {\it 23}, 193--217. 


\bibitem{Ko_2018}
Ko, B.; Song, J.W.; Chang, W. 
Crash forecasting in the Korean stock market based on the log-periodic structure and 
pattern recognition.
{\it Physica A} {\bf 2018}, {\it 492}, 308--323.

\bibitem{Chen_2018}
S. Chen; S. Zheng; H. Meersman.
Testing for the burst of bubbles in dry bulk shipping market using log periodic power law model. 
{\it Maritime Busin. Rev.} {\bf 2018}, {\it 3}, 128--144. 

\bibitem{Jhun_2018}
Jhun, J,; Palacios, P.; Weatherall, J.O.
Market crashes as critical phenomena? Explanation, idealization, and universality in econophysics.
{\it Synthese} {\bf 2018}, {\it 195}, 4477--4505. 

\bibitem{Dai_2018}
Dai, B.; Zhang, F.; Tarzia, D.; Ahn, K. 
Forecasting inancial crashes: Revisit to log-periodic power law.
{\it Complexity} {\bf 2018}, {\it 2018}, 4237471. 

\bibitem{Song_2020}
Song, R.; Shu, M.; Zhu, W. 
The 2020 global stock market crash: Endogenous or exogenous?
{\it Physica A} {\bf 2022}, {\bf 585}, 126425. 

\bibitem{Shu_2024}
Shu, M.; Song, R.  
Detection of financial bubbles using a log-periodic power law singularity (LPPLS) model.
{\it WIREs Comput. Statist.} {\bf 2024}, {\it 16}, e1632.

\bibitem{Oguzhan_2025}
Cepni, O.; Gupta, R.; Nel, J.; Nielsen, J. 
Monetary policy shocks and multi-scale positive and negative bubbles in an emerging country: 
The case of India.
{\it Financ. Innov.} {\bf 2025}, {\it 11}, 35. 



\bibitem{Yamakov}
Yamakov, V.; Milchev, A.; Foo, G.M.; R.B.; Pandey R.B.; Stauffer, D. 
Log-periodic oscillations for biased diffusion of a polymer chain in a porous medium.
{\it Eur. Phys. J. B} {\bf 1999}, {\it 9}, 659--667.

\bibitem{Sienkiewicz}
Sienkiewicz, J.; Fronczak, P.; Holyst, J.
Log-periodic oscillations due to discrete effects in complex networks.
{\it Phys. Rev. E} {\bf 2007}, {\it 75}, 066102.

\bibitem{Faillettaz}
Faillettaz, J.; Pralong, A.; Funk, M.; Deichmann, N. 
Evidence of log-periodic oscillations and increasing icequake activity during the breaking-off 
of large ice masses.
{\it J. Glaciol.} {\bf 2008}, {\it 54}, 725--737.  

\bibitem{Khamzin}
Khamzin, A.A.; Nigmatullin, R.R.; Popov, I.I.; Zhelifonov, M.P.
Log-periodic oscillations in the specific heat behaviour for self-similar Ising type spin systems.
{\it J. Phys. Conf. Ser.} {\bf 2012}, {\it 394}, 012008.

\bibitem{Bazak}    
Bazak, B.; Barnea, N.
Log-periodic oscillations in the photo response of Efimov trimers.
{\it Few Body Syst.} {\bf 2014}, {\it 55}, 851--856.

\bibitem{Thiem}
Thiem, S.
Origin of the log-periodic oscillations in the quantum dynamics of electrons in quasiperiodic 
systems.
{\it Phil. Mag.} {\bf 2015}, {\it 95}, 1233--1243.  

\bibitem{Rybczynski}
Rybczynski, M.; Wilk, G.; Wlodarczyk, Z.
System size dependence of the log-periodic oscillations of transverse momentum spectra.
{\it EPJ Web Conf.} {\bf 2015}, {\it 90}, 01002.

\bibitem{Wang}
Wang, H. et al.
Discovery of log-periodic oscillations in ultraquantum topological materials.
{\it Sci. Adv.} {\bf 2018}, {\it 4}, NO 11. 

\bibitem{Wang_2}
Wang, H. et al.
Log-periodic quantum magneto-oscillations and discrete-scale invariance in topological 
material HfTe$_5$.
{\it Nat. Sci. Rev.} {\bf 2019}, {\it 6}, 914--920.  

\bibitem{Bhoyar}
Bhoyar, P.D.; Gade, P.M.
Emergence of logarithmic-periodic oscillations in contact process with topological disorder.
{\it Phys. Rev. E} {\bf 2021}, {\it 103}, 022115.

\bibitem{Banerjee}
Banerjee, A.; Pavithran, I.; Sujith, R.I.
Imprints of log-periodicity in thermoacoustic systems close to lean blowout.
{\it Phys. Rev. E} {\bf 2023}, {\it 107}, 024219.

\bibitem{Dorbath}
Dorbath, E.; Gulzar, A.; Stock, G.
Log-periodic oscillations as real-time signatures of hierarchical dynamics in proteins.
{\it J. Chem. Phys.} {\bf 2024}, {\it 160}, 074103. 

\bibitem{Luck_2024}
Luck, J.M.
Revisiting log-periodic oscillations.
{\it Physica A} {\bf 2024}, {\it 643}, 129821. 



\bibitem{Yukalov_2021}
Yukalov, V.I.; Yukalova, E.P.
From asymptotic series to self-similar approximants.
{\it Physics} {\bf 2021}, {\it 3}, 829--878. 

\bibitem{Yukalov_PRD}
Yukalov, V.I.; Yukalova, E.P.
Self-similar extrapolation in quantum field theory.
{\it Phys. Rev. D} {\bf 2021}, {\it 103}, 076019. 

\bibitem{Yukalov_APNY}
Yukalov, V.I.; Yukalova, E.P.
Strong-coupling limits induced by weak-coupling expansions.
{\it Ann. Phys. (N.Y.)} {\bf 2024}, {\it 467}, 169716.


\bibitem{Mayer}
Mayer, J.E.; Mayer, M.
{\t Statistical Mechanics};
Wiley: London, 1977.

\bibitem{Feynman}
Feynman, R.P.
{\it Statistical Mechanics};
Benjamin: London, 1982.

\bibitem{Hansen}
Hansen, J.; McDonald, I.
{\it Theory of Simple Liquids};
Academic Press: Amsterdam, 1990.


\bibitem{Santos_1995}
Santos, A.; Lopez de Haro, M.; Bravo Yuste, S. 
An accurate and simple equation of state for hard disks.
{\it J. Chem. Phys.} {\bf 1995}, {\it 103}, 4622--4625.  

\bibitem{Clisbi_2006}
Clisby, N.; McCoy, B.M.
Ninth and tenth order virial coefficients for hard spheres in $D$ dimensions. 
{\it J. Stat. Phys.} {\bf 2006}, {\it 122}, 15--57. 

\bibitem{Mulero_2009}
Mulero, A.; Cachadina, I.;  Solana, J.R. 
The equation of state of the hard-disc fluid revisited. 
{\it Mol. Phys.} {\bf 2009}, {\it 107}, 1457--1465. 

\bibitem{Maestre_2011}
Maestre, M.A.G.; Santos, A.; Robles, M.; Lopez de Haro, M.
On the relation between coefficients and the close-packing of hard disks and hard spheres.
{\it J. Chem. Phys.} {\bf 2011}, {\it 134}, 084502. 



\bibitem{Yukalov_2003}
Yukalov, V.I.; Gluzman, S.; Sornette, D.
Summation of power series by self-similar factor approximants.
{\it Physica A} {\bf 2003}, {\it 328}, 409--438. 

\bibitem{Gluzman_2003}
Gluzman, S.; Yukalov, V.I.; Sornette, D.
Self-similar factor approximants.
{\it Phys. Rev. E} {\bf 2003}, {\it 67}, 026109. 

\bibitem{Yukalov_2022}
Yukalov, V.I.; Yukalova, E.P.
Self-similar sequence transformation for critical exponents.
{\it Phys. Lett. A} {\bf 2022}, {\it 425}, 127899.  

\bibitem{Niemeijer_1976}
Niemeijer, T.;  van Leeuwen, J.M.
Renormalization theory for Ising-like spin systems.
{\it Phase Transitions and Critical Phenomena}, vol. 6, pp. 425--505
Domb, C.; Green M.S. (eds.)
Academic, New York, 1976.

\bibitem{Gluzman_2002}
Gluzman, S.; Sornette, D.
Log-periodic route to fractal functions.
{\it Phys. Rev. E} {\bf 2002}, {\it 65}, 036142. 

\bibitem{Gell_23}
Gell-Mann, M.; Low, F.F.
Quantum electrodynamics at small distances.
{\it Phys. Rev.} {\bf 1954}, {\it 95}, 1300--1312. 

\bibitem{Bogolubov_24}
Bogolubov, N.N.; Shirkov,D.V.
{\it Introduction to the Theory of Quantized Fields};
Wiley: New York, 1980.

\bibitem{JohSoranti99}
Johansen, A.; Sornette, D.
Financial ``anti-bubbles'': Log-periodicity in gold and nikkei collapses.
{\it Int. J. Mod. Phys. C} {\bf 1999}, {\it 10}, 563--575. 

\bibitem{JohSorJan00}
Johansen, A.; Sornette, D.
Evaluation of the quantitative prediction of a trend reversal on the Japanese stock 
market in 1999. 
{\it Int. J. Mod. Phys. C} {\bf 2000}, {\it 11}, 359--364. 

\bibitem{SorZhou0002deep}
Sornette, D.; Zhou, W.-X.
The US 2000-2002 market descent: How much longer and deeper? 
{\it Quantit. Finance} {\bf 2002}, {\it 2}, 468--481. 

\bibitem{JohSorfts50}
Johansen, A.; Sornette, D.
Finite-time singularity in the dynamics of the world population and economic indices. 
{\it Physica A} {\bf 2001}, {\it 294}, 465--502. 

\bibitem{ZhouSoranti03}
Zhou, W.-X.; Sornette, D.
Testing the stability of the 2000-2003 US stock market ``antibubble''.
{\it Physica A} {\bf 2005}, {\it 348}, 428--452. 


\bibitem{Euler_1777}
Euler, L.
De formulis exponentialibus replicatis.
{\it Acta Acad. Petropolitanae} {\bf 1777}, {\it 1}, 38--60. 

\bibitem{Barrow}
Barrow, D.F.
Infinite Exponentials
{\it Am. Math. Monthly} {|bf 1936}, {\it 43}, 150--160.

\bibitem{Bell}
Bell, E.T.
The iterated exponential integers.
{\it Ann. Math.} {\bf 1938}, {\it 39}, 539--557.

\bibitem{Knoebel_1981}
Knoebel, R.A.
Exponentials reiterated.
{\it Am. Math. Monthly} {\bf 1981}, {\it 88}, 235--252.

\bibitem{Rippon}
Rippon, P.J.
Infinite exponentials.
{\it Math. Gazette} {\bf 1983}, {\it 67}, 189--196.

\bibitem{Bromer}
Bromer, N.
Superexponentiation. 
{\it Math. Mag.} {\bf 1987}, {\it 60}, 169--174.

\bibitem{Bender_1996}
Bender, C.M.; Vinson, J.P.
Summation of power series by continued exponentials.
{\it J. Math. Phys.} {\bf 1996}, {\it 37}, 4103--4120.

\bibitem{Anderson}
Anderson, J.
Iterated exponentials.
{\it Am. Math. Monthly} {\bf 2004}, {\it 111}, 668--679.

\bibitem{Devaney}
Devaney, R.L.; Josic, K.; Moreno Rocha, M.; Seal, P.; Shapiro, Y.; Frumosu, A.T.
Playing catch-up with iterated exponentials.
{\it Am. Math. Monthly} {\bf 2004}, {\it 111}, 704--709.

\bibitem{Hooshmand}
Hooshmand, M.H. 
Ultra power and ultra exponential functions. 
{\it Integr. Transf. Spec. Funct.} {\bf 2006}, {\it 17}, 549--558.

\bibitem{Marshall}
Marshall, A.J.; Tan, Y. 
A rational number of the form $a^a$ with $a$ irrational.
{\it Math. Gazette} {\bf 2012}, {\it 96}, 106--109.

\bibitem{Paulsen}
Paulsen, W. 
Tetration for complex bases. 
{\it Adv. Comput. Math.} {\bf 2018}, {\it 45}, 243--267.



\bibitem{Yukalov_25}
Yukalov, V.I.; Gluzman, S.
Self-similar exponential approximants.
{\it Phys. Rev. E} {\bf 1998}, {\it 58}, 1359--1382. 

\bibitem{Yukalov_2002}
Yukalov, V.I.; Yukalova, E.P.
Self-similar structures and fractal transforms in approximation theory.
{\it Chaos Solit. Fract.} {\bf 2002}, {\it 14}, 839--861. 

\bibitem{Gluzman_26}
Gluzman, S.; Sornette, D.; Yukalov, V.I.
Reconstructing generalized exponential laws by self-similar exponential approximants.
{\it Int. J. Mod. Phys. C} {\bf 2003}, {\it 14}, 509--527. 

\bibitem{Abhignan_2021}
Abhignan, V.; Sankaranarayanan, R.
Continued functions and perturbation series: Simple tools for convergence of
diverging series in $O(n)$-symmetric $\vp^4$ field theory at weak coupling limit.
{\it J. Stat. Phys.} {\bf 2021}, {\it 183}, 4. 

\bibitem{Abhignan_San_2021}
Abhignan, V.; Sankaranarayanan, R.
Continued functions and Borel-Leroy transformation: Resummation of six-loop
$\epsilon$-expansions from different universality classes.
{\it J. Phys. A} {\bf 2021}, {\it 54}, 425401. 

\bibitem{Abhignan_2023}
Abhignan, V.; Sankaranarayanan, R.
Continued functions and critical exponents: Tools for analytical continuation of
divergent expressions in phase transition studies.
{\it Eur. Phys. J. B} {\bf 2023}, {\it 96}, 31. 

\end{thebibliography}
\end{document}